\documentclass[a4paper,fleqn]{cas-dc}

\usepackage[numbers]{natbib}
\usepackage{graphicx}
\usepackage{caption}
\usepackage{subcaption}
\usepackage{svg}
\usepackage{booktabs} 
\usepackage{longtable}
\usepackage{multirow}
\usepackage{diagbox}

\usepackage[ruled,linesnumbered]{algorithm2e}

\def\tsc#1{\csdef{#1}{\textsc{\lowercase{#1}}\xspace}}
\tsc{WGM}
\tsc{QE}
\begin{document}
\let\WriteBookmarks\relax
\def\floatpagepagefraction{1}
\def\textpagefraction{.001}
\let\printorcid\relax % 可去掉页面下方的ORCID(s)

% \shorttitle{<short title of the paper for running head>} 
\shorttitle{MDF: A Dynamic Fusion Model for Multi-modal Fake News Detection}    

% \shortauthors{<short author list for running head>}
\shortauthors{Hongzhen Lv et al.}

\title[mode = title]{MDF: A Dynamic Fusion Model for Multi-modal Fake News Detection}  

% \tnotemark[<tnote number>] 
\tnotemark[1]

\tnotetext[1]{This work is a research achievement supported by the "Tianshan Talent" Research Project of Xinjiang (No.202304120002), the National Natural Science Foundation of China (No.02204120017), the Science and Technology Program of Xinjiang (No.202204120025), and the National Key R\&D Program of China Major Project (No.2022ZD0115800).}
% \tnotetext[2]{The second title footnote which is a longer text matter to fill through the whole text width and overflow into another line in the footnotes area of the first page.}
\author[label1,label2]{Hongzhen Lv}
\ead{107552304103@stu.xju.edu.cn}
\author[label1,label2]{Wenzhong Yang}
\ead{yangwenzhong@xju.edu.cn}\cormark[1]
\author[label1,label2]{Fuyuan Wei}
\ead{wfy@stu.xju.edu.cn}
\author[label1,label2]{Jiaren Peng}
\ead{107552201362@stu.xju.edu.cn}
\author[label1,label2]{Haokun Geng}
\ead{107552203997@stu.xju.edu.cn}
\address[label1]{School of information Science and Engineering, Xinjiang University, Urumqi 830046, China}
\address[label2]{Xinjiang Key Laboratory of Multilingual Information Technology, Xinjiang University, China}
\cortext[1]{Corresponding author}

\begin{abstract}
Fake news detection has received increasing attention from researchers in recent years, especially multi-modal fake news detection containing both text and images.However, many previous works have fed two modal features, text and image, into a binary classifier after a simple concatenation or attention mechanism, in which the features contain a large amount of noise inherent in the data, which in turn leads to intra- and inter-modal uncertainty.In addition, although many methods based on simply splicing two modalities have achieved more prominent results, these methods ignore the drawback of holding fixed weights across modalities, which would lead to some features with higher impact factors being ignored.To alleviate the above problems, we propose a new dynamic fusion framework dubbed \textbf{MDF} for fake news detection.As far as we know, it is the first attempt of dynamic fusion framework in the field of fake news detection.Specifically, our model consists of two main components:(1) \textbf{UEM} as an uncertainty modeling module employing a multi-head attention mechanism to model intra-modal uncertainty; and (2) \textbf{DFN} is a dynamic fusion module based on D-S evidence theory for dynamically fusing the weights of two modalities, text and image.In order to present better results for the dynamic fusion framework, we use GAT for inter-modal uncertainty and weight modeling before DFN.Extensive experiments on two benchmark datasets demonstrate the effectiveness and superior performance of the MDF framework.We also conducted a systematic ablation study to gain insight into our motivation and architectural design.We make our model publicly available to:\url{https://github.com/CoisiniStar/MDF}
\end{abstract}

\maketitle

% Main text
%-------------------------------------------------------------------------------- 1
\section{Introduction}
\label{introduction}

With the development of the Internet, especially the popularization of multimedia, multi-modal posts with text and images have gradually become popular on social media such as Twitter and Weibo.More and more people share what they see and hear through social platforms, and social media can tweet users about social events happening around the world in a timely manner.However, it also contains many tweets with fake news at the same time.In recent decades, major emergencies have occurred frequently, and fake news has been proliferating in our lives, which has seriously triggered social panic.For example, in the 2016 U.S.presidential election\cite{Survey1}, the proliferation of fake news on social media seriously affected citizens' intention to vote.The fairness of the election process as well as the result was equally negatively affected by fake news.Not only that, during the COVID-19 epidemic\cite{COVID-19}, social media was flooded with intentionally fabricated misinformation not limited to the fact that salt water, tea and vinegar can eliminate COVID-19.

\begin{figure}
    \centering
    \includegraphics[width=1\columnwidth]{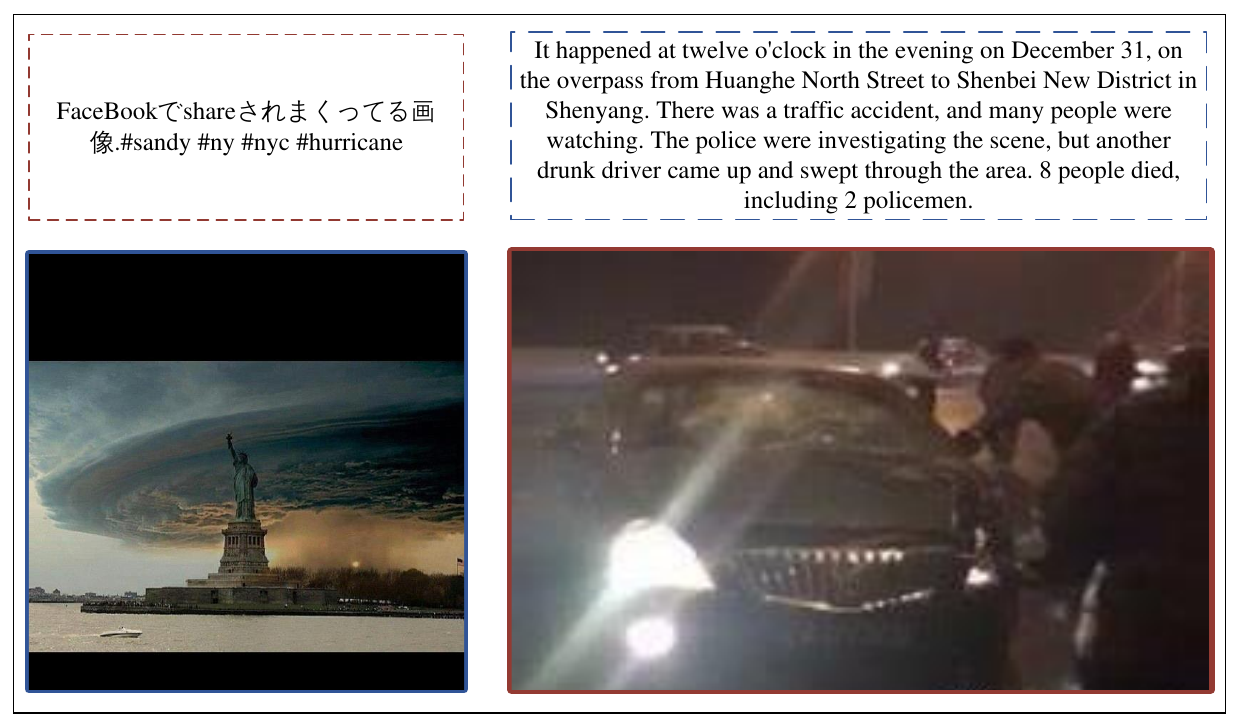}
    \caption{Sample low-quality multi-modal posts.Irregular news tweets are applied in the first one.The attached image with low resolution is shown in the second one.}
    \label{fig:intro}
\end{figure}

With the emergence of multi-modal news containing text and images, the task of fake news detection has become more and more challenging, and some researchers have begun to focus on multi-modal fake news detection.Vinyals et al.\cite{vinyals2015show} first proposed multi-modal fusion, which generates natural sentences to describe the images.Jin et al.\cite{withRNN} built an end-to-end network, where the fake news model was designed using RNN,while RNN utilizes local attention mechanism to combine text images and social context features.In order to improve the detection target of fake news, Qu et al.\cite{QMFND} used a quantum multi-modal fusion strategy to enhance the performance of fake news detection.Wu et al.\cite{MFIR} focused on inconsistent inference of news content to further improve the performance of the model.Although previous works obtained prospective performance, they still suffer from the following problems:
\begin{enumerate}
    \item Multimodal data come from heterogeneous sources\cite{Survey2}, which are of low quality and contain noise\cite{GaoLCZ20,ZhangWZHFZP23}.In particular, multimodal posts from social media, such as the two examples in the Fig.\ref{fig:intro}, the news texts in the first example are all non-normalized representations, which to some extent will cause some impact on the classification results; the news in the second example contains a low-resolution attached image, which will result in the obtained features containing non-excludable noise.However, previous research methods seldom consider the data uncertainty caused by the inherent noise of each modality, and only perform a simple static fusion strategy, which will lead to the superposition of noise between modalities and reduce the generalization performance of the model.
    \item Some traditional methods\cite{SAFE,MVAE,VQA}, perform a simple fusion based on the feature vectors acquired by the text encoder and the visual encoder, and feed the fused features into the classifier.As for attention-based methods\cite{withRNN,ZhangFQX19,WuZZWX21}, they learn the corresponding attention vectors based on specific modalities to balance the value of each modality's contribution to the final classification effect.However, these methods place the final feature vector in a high dimension, increasing the computational complexity of the model.The MM-ULN model\cite{WeiHZH23} takes into account the effect of intra- and inter-modal uncertainty on the final decision value, but it only uses the traditional approach of modal complementary features, where each modality contains a fixed weight, and is unable to give a high confidence level to the features with high impact factors.
\end{enumerate}

In order to better fuse post and accompanying image information from heterogeneous sources, we propose a dynamic fusion framework dubbed MDF for multi-modal fake news detection, in which the intra-modal uncertainty information is first modeled by an uncertainty estimation module named UEM.Subsequently,the representation features that form the robustness are fed into the Graph Attention Network(GAT) for inter-modal uncertainty modeling and learning the dynamic weights for each modality.Based on Dempster-Shafer evidence theory is applied before the final decision-making layer, which will decide the final strategy based on the dynamic weight values mentioned above.

We have experimented the MDF framework on two public datasets, Twitter and Weibo,the experimental results show that our model exhibits excellent performance in the multi-modal fake news detection task.Recent advances in Large Language Modeling (LLMs) have excellent performance in various NLP endeavors, but they still face great challenges in multimodal fake news detection tasks. For example, it is mentioned in [13] that although complex LLMs like GRT3.5 can provide desirable multi-perspective justifications for exposing fake news, they are still not as effective as basic SLMs (Small Language Models) such as fine-tuned BERT.Therefore, we focus on fine-tuning our small model specifically for multimodal fake news detection.

Our main contributions can be summarized as follows:

$\bullet$ In order to better extract robust feature representations for multi-modal news content containing noise, we developed a multi-modal attention-based uncertainty modeling module,dubbed UEM,through the use of an attention-based strategy.

$\bullet$ We designed a multi-modal dynamic fusion module,named DFN, based on Dempster-Shafer evidence theory, which dynamically balances the contribution of both modalities to the final classification, and solves the drawbacks of weight fixation in the static fusion strategy in an effective way.

$\bullet$ Experimental results on two real-world multi-modal benchmark datasets demonstrate the effectiveness and superiority of our model.

The rest of this paper is organized as follows:First, we review the related work on fake news detection and the methods involved in Section \ref{sec:Related Work}.Second, the MDF framework is formalized in Section \ref{sec:Methodology} and the detailed design ideas of each layer are elicited.In addition, the experimental setup and results will be presented in Section \ref{sec:EXPERIMENTS}.The shortcomings of the work are also pointed out in Section \ref{sec:limitation}.In Section \ref{sec:conclusion},we summarize our main work and outlines the direction of future work.
%-------------------------------------------------------------------------------- 2
\section{Related Work}
\label{sec:Related Work}
In this section, we briefly review previous research results related to multi-modal fake news detection.In addition, relevant research results on multi-modal dynamic fusion strategies as well as Dempster-Shafer evidence theory, which are techniques related to our approach, are reviewed.
%--------------------------------------------------------------------------------2.1
\subsection{Multi-modal Fake News Detection}
As multi-modal continues to evolve, several researchers have become enthusiastic about text and image based task processing.Antol et al.\cite{VQA} proposed a visual question and answer task to learn the representation of features from multiple aspects.This work promotes cross research in the field of image understanding and natural language processing and also lays the foundation for multi-modal fake news detection task.EANN (Event Adversarial Neural Network) model is proposed in the work of Wang et al.\cite{Eann} which uses an event learner to learn feature representations of text and images in an article, but the additional auxiliary features also increase the cost of detection.MVAE\cite{MVAE} extracts intra-modal information from two unimodal modalities, text and image, respectively.And it uses a multi-modal variant encoder for simple fusion.SpotFake\cite{SpotFake} uses a pre-trained language model to extract features.Intuitively, it uses large-scale textual and visual pre-trained language models to extract features for each unimodal state and simply concatenates them to form a multi-modal representation vector, which is then fed into a binary classifier to detect fake news.In much the same way as previous approaches, the SAFE\cite{SAFE} model also uses simple fusion to detect fake news in a similarity-based manner.Wu et al.\cite{WuZZWX21} proposed Multi-modal Common Attention Network (MCAN) for fake news detection, which learns the inter-dependencies between multi-modal features and achieves good results in the field of fake news detection.Wang et al.\cite{MetaNeural} provided a fake news detection framework called MetaFEND,which can be effective in detecting breaking news,because it learns from a small number of verified posts.Tuan et al.\cite{withBERT-Att} in their work showed a new approach to detect fake news by effectively learning and fusing multi-modal features in posts.Song et al.\cite{Knowledge-augmented} added Graph Convolutional Networks(GCN) to their framework called KMGCN by using fused textual information with knowledge concepts and visual information for multi-modal modal semantic representation.In addition, some other researchers, e.g., Xue et al.\cite{Byconsistency}, Fung et al.\cite{fung2021infosurgeon}, explored the consistency of multi-modal data to help detect fake news.In the above approach, the researchers did not pay attention to the drawbacks of fixed weights, so in order to improve the problem, we propose a dynamic fusion mechanism based on Dempster-Shafer thereby realizing the assignment of a confidence score to each modality, which results in the dynamic change of the weights of each modality.
%--------------------------------------------------------------------------------2.2

\subsection{Multi-modal Dynamic Fusion}
Multi-modal dynamic fusion facilitates the model to take full advantage of the complementary features of different modalities, which in turn improves the robustness and generalization performance.In order to better analyze human multi-modal languages, Han et al.\cite{han2022multimodal} proposed a new trustworthy multi-modal classification algorithm that dynamically evaluates both feature and modality levels of information for different samples, thus credibly integrating multiple modalities.Zhang et al.\cite{zhang2023provable}, based on theoretical analyses, further revealed that the generalization ability of dynamic fusion is consistent with the performance of uncertainty estimation, and proposed a generalized dynamic multi-modal fusion framework for low-quality data, which uses the generalization error obtained from multi-modal fusion to dynamically update each unimodal predictor, so that the multi-modal decision tends to rely more on the high-quality modes than on the other modes.The effect of unreliable modes is mitigated by dynamically determining the fusion weights of each modality.Inspired by the above studies, we introduce the dynamic fusion mechanism into the fake news detection task to dynamically balance the weights of each modality.

%--------------------------------------------------------------------------------2.3
\subsection{Dempster-Shafer evidence theory}
Dempster-Shafer evidence theory, also known as confidence function theory, is a complete theoretical architecture that integrates multi-source information and is widely used in multi-modal fusion tasks.D-S evidence theory was first introduced in \cite{yager2008classic}.After this,\cite{DTBM} introduced a mathematical evidence theory and applied it to multi-source information fusion.Currently, DS evidence theory has a very wide range of applications in the field of fusion decision making\cite{denoeux2019decision}.The D-S evidence theory framework facilitates the development of Bayesian theory, and at the same time relaxes the limitations of the Bayesian theoretical reasoning methods\cite{Survey3}.It aims to compute and update the confidence of a proposition by merging evidence from multiple sources, and it has also become a mathematical framework for dealing with uncertainty.However, traditional synthesis rules cannot be directly applied in the MDF framework, so we will elaborate our algorithm in Section \ref{sec:DFN}.
We feed robust representational features formed by an uncertainty modeling module composed of multi-head attention into the fusion network to learn dynamic inter-modal weight modeling.Finally, we use Dempster-Shafer evidence theory in order to achieve dynamic fusion of two modalities, text and image, providing a dynamic fusion framework for multi-modal fake news detection.

\begin{figure*}
    \centering
    \includegraphics[width=2\columnwidth]{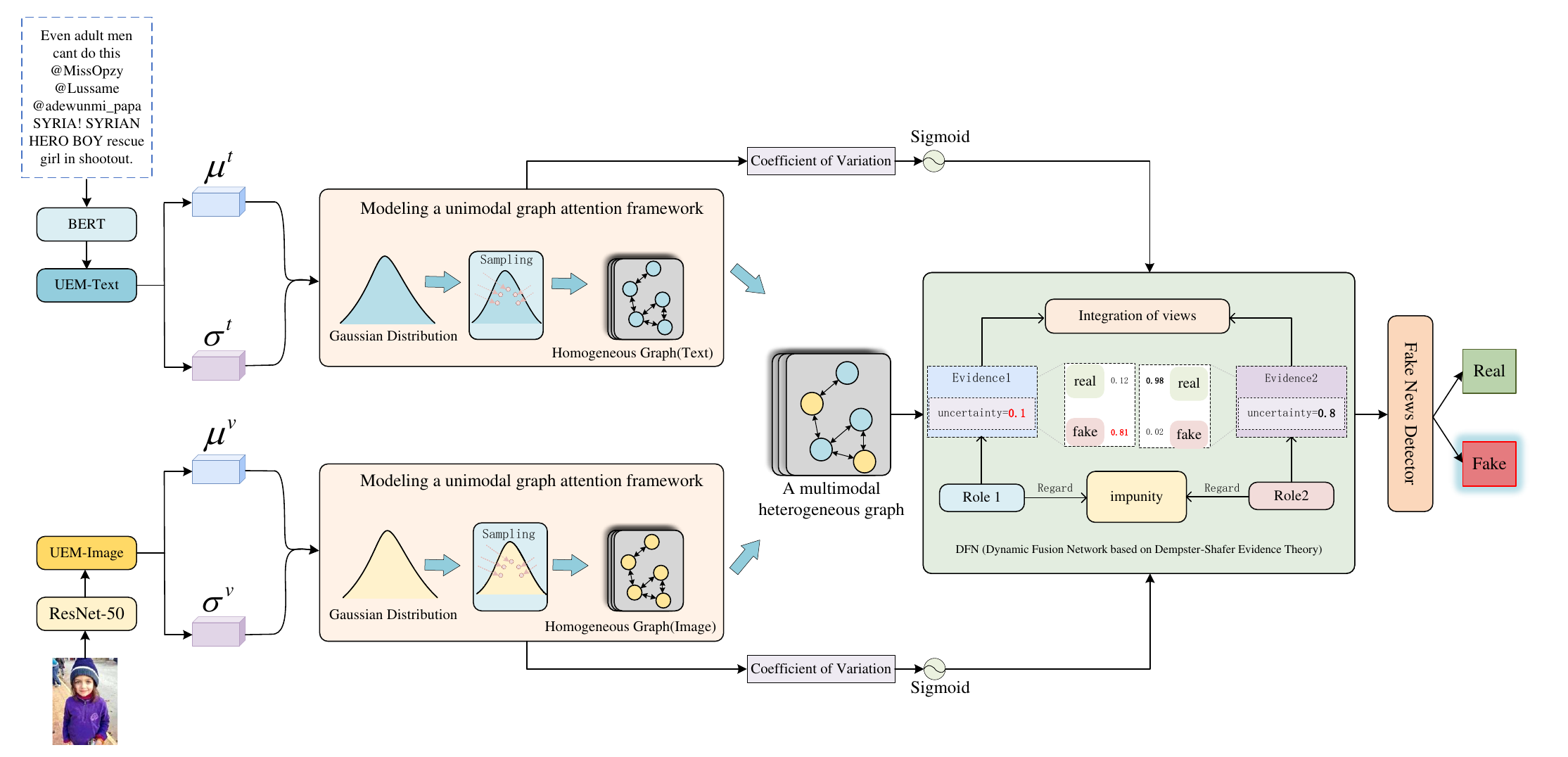}
    \caption{Overview of MDF:It mainly consists of a UEM module that employs an attention mechanism, a graph attention network, a DFN module that incorporates the Dempster-Shafer theory of evidence, and a fake news detector.Given a noisy post and accompanying image captured by a social platform, the UEM maps the tweet and the accompanying image into the corresponding potential subspaces, respectively, to complete the unimodal intra-modal uncertainty modeling. Subsequently, it is fed into GAT for inter-modal uncertainty modeling and two-modal weight modeling,the dynamic weight perception of the two modalities is completed using the DFN module containing Dempster-Shafer evidence theory.The final confidence of each modality is fed back to the fake news detector to complete the final dynamic fusion strategy.
}
    \label{fig:mdfmodel}
\end{figure*}

%-------------------------------------------------------------------------------- 3
\section{Methodology}
\label{sec:Methodology}

In this section, we detail a dynamic fusion framework for fake news detection based on D-S evidence theory and the Graph Attention Network(GAT).First in Section \ref{sec:overview}, we outline the proposed model,i.e.,a dynamic fusion framework for multi-modal fake news detection, named MDF.the uncertainty modeling module employing the attention mechanism is described in detail in Section \ref{sec:UEM}.Finally, a dynamic fusion module based on graph attention networks and Dempster-Shafer evidence theory is elicited in Section \ref{sec:DFN}.

%-------------------------------------------------------------------------------- 3,1
\subsection{Overview}
\label{sec:overview}
The dynamic fusion architecture MDF framework for multi-modal fake news detection is shown in Fig.\ref{fig:mdfmodel}.In order to model multi-modal news containing noise, we first feed semantic features formed by pre-trained language models (PLMs) into the UEM module employing the attention mechanism, which maps the semantic features into a latent subspace of a Gaussian distribution.i.e., the traditional point embeddings are represented in a latent Gaussian subspace containing mean $\mu$ and varianc $\sigma^2$.Subsequently, we model the inter-modal uncertainty using a graph attention network in the DFN module, and the Dempster-Shafer theory of evidence is used in the DFN for assigning a confidence score to the features of each modality and feeding that score back into the fake news detector.

%-------------------------------------------------------------------------------- 3,2
\subsection{Modeling Unimodal Uncertainty Using the Multi-head Attention Mechanism}

\label{sec:UEM}
\begin{figure}
    \centering
    \includegraphics[width=0.85\columnwidth]{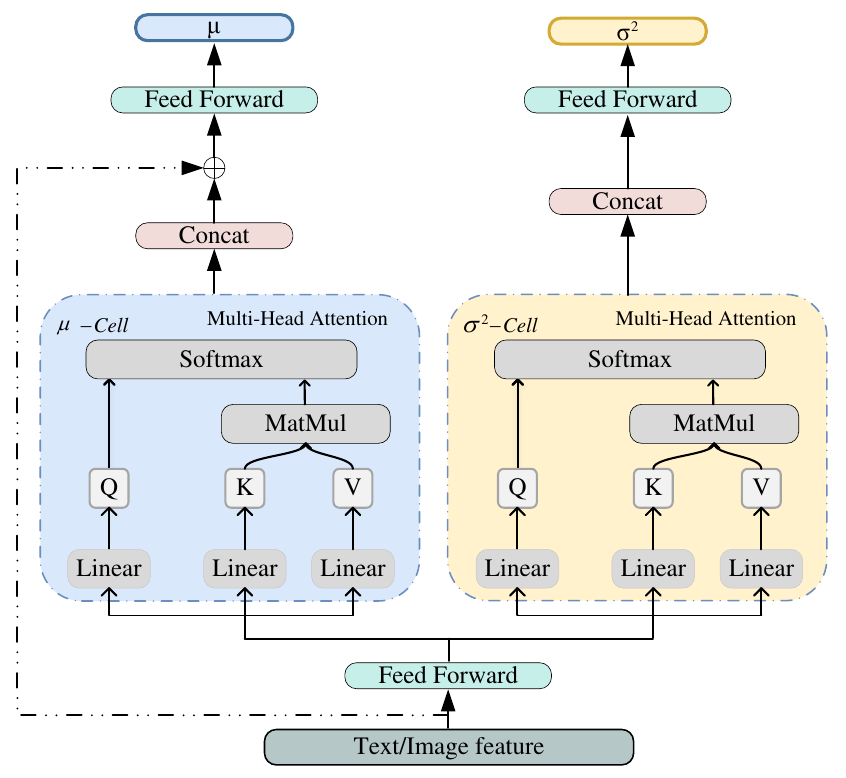}
    \caption{Diagram of the UEM architecture.The UEM architecture based on the multi-attention mechanism will represent each modality as a Gaussian distribution satisfying a mean of $\mu$ and a variance of $\sigma^2$ based on the noise inherent in its features.And the learned mean values are combined with the original unimodal features to form a robust representation of each modality.}
    \label{fig:UEM}
\end{figure}
Most social media posts exist in a multi-modal form (containing tweets and their accompanying images).We first use a pre-trained language model to learn the token-level embedding representations of posts used to extract linguistic features.Specifically, let $T=([CLS],t_1,t_2,...,t_n,[SEP])$ be a sequence of post texts with $n$ tokens,and $e^t=(e_1,...,e_n)$is an embedded representation of the context token level generated by the pre-trained BERT\cite{devlin2018bert},where $e^t\in R^{n\times d}$ and each token $e_i \in \mathcal{R}^d$ is a d-dimensional vector.Similarly, we follow many previous research works\cite{Eann}\cite{MVAE}\cite{WeiHZH23} and use the popular pre-trained backbone network (i.e., pre-trained Resnet-50) to extract region of interest (RoI) pooling features from post attachments to obtain fine-grained object-aware representations.Formally,$e^v = \{e^v_1,e^v_2,...,e^v_l\},e^v\in \mathbb{R}^{l\times d_v}$ where $e^v_i \in \mathbb{R}^{d_v}$ denotes the $i-th$ ROI feature, and i is the number of extracted ROI features.However, since the images accompanying user tweets are usually low-resolution, most of the textual content published by users is non-standardized content, resulting in a large amount of noise in multi-modal news.In order to model noise-containing images and text-formatted posts, we use the uncertainty modeling module of the multi-head attention mechanism to model the uncertainty of the two modalities.Specifically, we use a multivariate Gaussian distribution to model the extracted features of the two modalities, text and image, as a certain Gaussian distribution, and reconstruct the embedding representation of each modality using distribution representation sampling.In detail, we assume that the noise of each modality satisfies a certain Gaussian distribution, and for the input features of each modality, the UEM module computes the mean vector $\mu$ and the variance vector $\sigma^2$ that the noise satisfies, the mean vector represents the location of the Gaussian distribution in the probability space, and the variance vector represents the range in each of the dimensions, and its architecture is shown in Fig.\ref{fig:UEM}.In the multi-head attention operation, the input representation $H\in\mathbb{R}^{T\times D}$ is divided into $k$ heads, where $T$ means the unimodal feature length and $D$ indicates the size of the hidden layer.Two channels are used in the UEM module to represent the mean $\mu$ and variance $\sigma^2$ of the generated Gaussian distribution.Since we are using a multi-head attention mechanism, the input representation $H^{(i)}\in R^{T\times D/2k}$ is projected into the query vectors $Q^{(i)}$,$K^{(i)}$,$V^{(i)}$,where $i$ indicates the number of heads.
Formally,the multi-head attention used to generate the mean $\mu$ satisfied by unimodal textual noise can be indicated as:
{\small
\begin{equation}
        [Q^{i}_{\mu}(T),K^{i}_{\mu}(T),V^{i}_{\mu}(T)]=[W_q(T),W_k(T),W_v(T)]^T
        \times H^{(i)}_\mu(T)
        \label{eq:equation1}
\end{equation}
}
% {\small
% \begin{equation}
%        [Q^{i}_{\sigma^2}(T),K^{i}_{\sigma^2}(T),V^{i}_{\sigma^2}(T)]=[W_q(T),W_k(T),W_v(T)]^T\times H^{(i)}_{\sigma^2}(T)
%         \label{eq:equation2}
% \end{equation}
% }
The multi-head attention used to generate the mean $\mu$ satisfied by unimodal image noise can be indicated as:
{\small
\begin{equation}
        [Q^{i}_{\mu}(V),K^{i}_{\mu}(V),V^{i}_{\mu}(V)]=[W_q(T),W_k(V),W_v(V)]^T\times H^{(i)}_\mu(V)
        \label{eq:equation2}
\end{equation}
}
% {\small
% \begin{equation}
%         [Q^{i}_{\sigma}(V),K^{i}_{\sigma}(V),V^{i}_{\sigma}(V)]=[W_q(T),W_k(V),W_v(V)]^T\times H^{(i)}_\sigma(V)
%         \label{eq:equation4}
% \end{equation}
% }
We specify that $d_k$ denotes the value of $D/(2k)$.In the above equation,$W_q(T)\in \mathbb{R}^{d_k\times 3 d_k}$,$W_k(T)\in \mathbb{R}^{d_k\times 3 d_k}$ and $W_v(T)\in \mathbb{R}^{d_k\times 3 d_k}$ are the weighting matrices used to project the features into each header subspace with a weight matrix.$H^{(i)}_\mu(T)$ and $H^{(i)}_\mu(V)$ imply unimodal input features at the $i-th$ attention head for text and image, respectively.The $Q^{i}_{\mu}(T)$,$K^{i}_{\mu}(T)$,$V^{i}_{\mu}(T)$ indicate the query, key, and value vectors formed based on the textual features.Similarly, $Q^{i}_{\mu}(V)$,$K^{i}_{\mu}(V)$ and $V^{i}_{\mu}(V)$ indicate the query, key and value vectors formed based on image features.The attention vector for the variance $\sigma^2$ is similar to the mean and is described below only in terms of the mean.Furthermore, we follow \cite{vaswani2017attention} and for each attention head a strategy of scaling dot product attention is employed.e.g., the following Eq.\ref{eq:equation3} indicates the output of the $i-th$ head used to generate the unimodal text noise mean $\mu$:

\begin{equation}
        H^{(i)}=Action(\frac{Q^{(i)}_{\mu}(T))K^{(i)}_{\mu}(T)}{\sqrt{d_k}})\cdot V^{(i)}_{\mu}(T)
        \label{eq:equation3}
\end{equation}

where the term $Action$ denotes the activation and normalization functions.The $Q^{i}_{\mu}(T)$,$K^{i}_{\mu}(T)$ and $V^{i}_{\mu}(T)$ indicate the query, key, and value vectors for generating the mean value of the unimodal textual noise $\mu$ formed by the $i-th$ attention head.At the end, the vectors from the plurality of attention heads for representing the mean value $\mu$ of the noise present in the unimodal text modality are concatenated to form a robust representation of the unimodal text:
\begin{equation}
        MH^{Total}_{\mu}(T)=W_{out}(T)\cdot concat_{i\in [k]}[H^{(i)}_\mu(T)]
        \label{eq:equation4}
\end{equation}
where $concat$ instructs that each attention head $H^(i)$ from the mean satisfied by the modal noise used to generate a single text or single image be subjected to a concatenation operation along a certain dimension.A learnable weight matrix $W_{out}$ is introduced into the above equation to linearly transform the features after joining each attention head.
For the final unimodal characterization we used the sampling reparameterization strategy, but since the sampling approach is a non-differentiable operation, also the sampling process presents challenges in terms of suppressing the gradient back propagation.Thus inspired by \cite{WeiHZH23}, the reparameterization trick is introduced into our architecture,i.e.,we sample a random variable from a standard normal distribution instead of performing the sampling operation directly in $\mathcal{N}(\mu,\sigma^2)$:

\begin{equation}
        \centering
        z=\mu+\varepsilon\sigma
        \label{eq:equation5}
\end{equation}

After Eq.\ref{eq:equation5}, we output $z$ based on the predictive distribution derived from the UEM.Thus, we separate the computation of the mean and standard deviation from the sampling operation, and this decoupling operation turns the original parameters into trainable ones.
After the UEM module, we get a more robust representation feature for each modality because it contains the distribution satisfied by each modal noise than the traditional point embedding approach.This robustness feature is excellent for detecting multi-modal fake news, which we will elaborate later in the ablation experiments.
%-------------------------------------------------------------------------------- 3,3
\subsection{A Dynamic Fusion Module Based on Graph Attention Network and Dempster-Shafer Evidence Theory}
\label{sec:DFN}

The simple concatenate mechanism has been applied to many previous multi-modal fake news detection tasks\cite{SAFE} \cite{MVAE} \cite{SpotFake}, where the text and image unimodalities hold fixed fusion weights and some of the posts have high uncertainty due to a variety of reasons, so the DFN module is designed and the confidence of each module is taken into account by it in its internal structure.Specifically, we model the unimodal feature representation satisfying a certain Gaussian distribution reconstructed by the UEM module as a graph structure.And we perform graph attention operations in two unimodal graph structures:

\begin{equation}
        G_t^{m^l}=\displaystyle\sum_{j\in \mathcal{N}(i)}\alpha_{i,j}W^l_tm^{l-1}_j
    \label{eq:equation6}
\end{equation}
\begin{equation}
        G_v^{m^l}=\displaystyle\sum_{j\in \mathcal{N}(i)}\alpha_{i,j}W^l_vm^{l-1}_j
    \label{eq:equation7}
\end{equation}

In the Eq.\ref{eq:equation7},$m^l$ indicates the node embedding update representation of the $l-th$ layer.$W^l$ indicates the weight parameter of the $l-th$ layer.$\alpha_{i,j}$ denotes the attention score between node $i$ and node $j$ in the unimodal graph,The exact representation is shown in the following Eq.\ref{eq:equation8} and Eq.\ref{eq:equation9}:

\begin{equation}
        a^l_{i,j}=Softmax(e^l_{i,j})
    \label{eq:equation8}
\end{equation}

\begin{equation}
        e^l_{i,j}=LeakyReLU( \vec{\alpha} [Wm_i||Wm_j])
    \label{eq:equation9}
\end{equation}

where $||$ is a splicing operation and LeakyReLU is an abbreviation for Leaky Rectified Linear Unit.
To model inter-modal uncertainty, we incorporate a heterogeneous graph structure $G_{mm}$ after two reconstructed unimodal features.For $G_{mm}$,we still use graph attention mechanism.In addition, to obtain confidence scores for each modality, we feed the reconstructed features obtained by UEM into the Dempster-Shafer evidence theory module.However, the traditional D-S fusion approach is not suitable for the MDF framework, so we reconstruct the new fusion approach.Specifically, we will first calculate the confidence score for each modality based on the normalized coefficient of variation\cite{li2023zico} defined in  Eq.\ref{eq:equation10} and Eq.\ref{eq:equation11}.

\begin{equation}
       {Score}^{T} = sigmoid(\frac{\sqrt{\sigma_t}}{\mu_t}*scal)
    \label{eq:equation10}
\end{equation}

\begin{equation}
        {Score}^{V} = sigmoid(\frac{\sqrt{\sigma_v}}{\mu_v}*scal)
    \label{eq:equation11}
\end{equation}

\begin{figure}[!t]
    \centering
    \includegraphics[width=0.5\columnwidth]{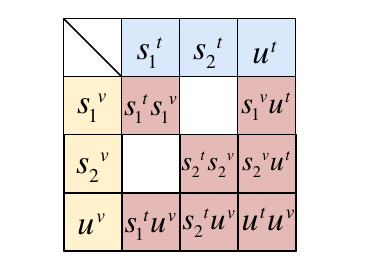}
    \caption{Diagram of the effect of binary classification based on D-S evidence theory. The light blue region is the confidence score derived from the single text modality, and the light yellow region is the confidence score derived from the single image modality. The white region represents the conflict region of the two decisions. The light red region represents the part where the decision values of the two modalities are compatible.}
    \label{fig:DST}
\end{figure}

In the above equation,$\mu^t$ and $\mu^v$ denote the reconstructed mean of the text and image respectively,$\sigma_t$ and $\sigma_v$ denote the reconstructed variance of the text and image, respectively.$scal$ is a scaling factor,$sigmoid$ is an activation function that maps the input to [0,1].
In this way, the dynamic weights of the two modalities with respect to the final decision value are captured.Subsequently, we combine the confidence scores (normalized coefficients of variation) of each modality with respect to the final decision value and the uncertainty inherent in each modality to arrive at the final decision opinion,as shown in Fig.\ref{fig:DST}.We take full advantage of the Dempster-Shafer evidence-theoretic fusion: when the two modal confidence levels are low, the final classification must belong to the low confidence set, at which time we will utilize the powerful advantage of the multi-modal heterogeneous graph to arrive at the final decision value; when the two modal confidence levels are high, the final classification result will generally be a subset of the high confidence set, at which time we randomly choose the decision value of one modality.When and only one modal confidence is high, the final classification is determined by that modal; when two modal decisions conflict, the final classification confidence is reduced, at this time, the decision-making effect of the multi-modal heterogeneous graph is adopted.Specifically, we will set a threshold  as shown in Algorithum \ref{alg:alg1} as a measure, and the choice of this threshold is also discussed in detail in Section \ref{sec:Analysis Dfn Params}.

\begin{algorithm*}
    
    \SetAlgoLined %显示end
	\caption{Dempster-Shafer evidence theory algorithm for multimodal fake news detection}%算法名字
    \label{alg:alg1}
	\KwIn{Decision values in two unimodal perspective $\mathcal{D}_{t}$,$\mathcal{D}_{v}$ and their uncertainty scores $\mathcal{S}_t$ and $\mathcal{S}_v$}  %输入参数
        
	\KwOut{High confidence decision-making strategies $\mathcal{P}$}%输出

        Normalized decision values for unimodal text states:
        \\
        \hspace{4mm}$\mathcal{N}_{t}=\frac{1}{1+exp(-\mathcal{D}_t)}$,$\mathcal{N}_{v}=\frac{1}{1+exp(-\mathcal{D}_v)}$

        Get two unimodal decision strategies for the object:
        \\
        \hspace{4mm}$m1[\mathcal{A}]=\mathcal{D}_{t_{out1}}$,$m1[\mathcal{B}]=\mathcal{D}_{t_{out2}}$
        
        \hspace{4mm}$m2[\mathcal{A}]=\mathcal{D}_{v_{out1}}$,$m2[\mathcal{B}]=\mathcal{D}_{v_{out2}}$
        \\
        Fusion of evidence obtained from two modalities using uncertainty scores:
        \\
        \hspace{4mm}$m[\mathcal{A}] = m1[\mathcal{A}] \ast m2[\mathcal{A}] + m1[\mathcal{A}] \ast  \mathcal{D}_v + \mathcal{D}_t \ast m2[\mathcal{A}]$
        \\
        \hspace{4mm}$m[\mathcal{B}] = m1[\mathcal{B}] \ast m2[\mathcal{B}] + m1[\mathcal{B}] \ast  \mathcal{D}_v + \mathcal{D}_t \ast m2[\mathcal{B}]$
        \\
        Get the uncertainty values of the fused two modes:
        \\
        \hspace{4mm}$m[\mathcal{\mathcal{A}}\bigcup \mathcal{B}] = \mathcal{D}_t \ast \mathcal{D}_v$
        \\
        Determine whether the overall uncertainty is greater than the threshold $\gamma$:

		\If{$m[\mathcal{\mathcal{A}}\bigcup \mathcal{B}]$ > $\gamma$}{
		  Decision values obtained using the theory of evidence are not credible:
            \\
            \hspace{4mm}$uncertainty = True$
		}
        \Else{
            Decision values obtained using the theory of evidence are credible:
            \\
            \hspace{4mm}$uncertainty = False$
        }
		\If{$uncertainty = False$ and $\mathcal{D}_t > \mathcal{D}_v$}{
            The output of unimodal text should be used as a metric:
            \\
            $\mathcal{P}=max(m1[\mathcal{A}],m1[\mathcal{B}])$
		}
        \ElseIf {$uncertainty = False$ and $\mathcal{D}_t < \mathcal{D}_v$}{
            The output of unimodal image should be used as a metric:
            \\
            $\mathcal{P}=max(m2[\mathcal{A}],m2[\mathcal{B}])$
        }
        \Else{
            The output of the multimodal graph attention mechanism is used as a metric:
            $\mathcal{P}=\mathcal{G}_m$
        } 
        \KwRet $\mathcal{P}$
\end{algorithm*}

%--------------------------------------------------------------------------------3.4
\subsection{Fake News Detector}
\label{sec:Detector}
We follow previous research work \cite{SAFE}\cite{MVAE}\cite{ZhangFQX19}, among others, and at the end of the Dempster-Shafer evidence-theoretic fusion framework, we use a linear neural network with an activation function to classify the robustness features of the resulting multi-modal fake news detection.Formally, it can be formulated as:

\begin{equation}
       \hat{y}_n=Softmax(W_nh+b_n)
    \label{eq:equation12}
\end{equation}

In the above equation,$W_n$ and $b_n$ are trainable parameters.In order to prevent the bad overfitting phenomenon from occurring, we add the Dropout mechanism after the linear layer.

%--------------------------------------------------------------------------------3.5
\subsection{Optimization Algorithm}
\label{sec:OptimizationAlgorithm}
We tried to optimize our classification model using a loss function as defined in Eq.\ref{eq:equation13}, which is shown below:

\begin{equation}
    \mathcal{L}(\theta)=\frac{1}{N}Y_ilog\hat{P}^0_i+(1-Y_i)log(1-\hat{P}^1_i)
    \label{eq:equation13}
\end{equation}

Note that N indicates the number of all news samples.
However using only $\mathcal{L}(\theta)$ as the final loss function will result in variance collapse.Since all sample point vectors would converge to the optimum, the final distributional representation would degenerate into a deterministic point representation, which would cause the model to lose its ability to learn multi-modal uncertainty.To prevent this bad phenomenon from happening, we introduced a penalty term, $\mathcal{L}_{reg}$ , which is used to prevent the distribution's level of uncertainty from falling below the exact lower bound $eta$.Therefore, we introduce the entropy-defined penalty term of the Gaussian distribution into our training framework to realize the above operation, formally redefining the loss function formulation as:
\begin{equation}
    \mathcal{L}_{pred}=\mathcal{L}(\theta)+\alpha\mathcal{L}_{reg}
    \label{eq:equation14}
\end{equation}

where $\mathcal{L}(\theta)$ is the cross-entropy loss function defined by Eq.\ref{eq:equation15}.Formally, $\mathcal{L}_{reg}$ can be expressed as:

\begin{equation}
    \mathcal{L}_{reg}=max(0,\eta-h(\mathcal{N}(\mu,\sigma^2)))
    \label{eq:equation15}
\end{equation}

The $\eta$ in the above equation is a lower bound value, which we have taken to be 0.01 to control the minimum level of uncertainty in the final distribution.In subsection 4.7.3, we will discuss in detail the role of this hyperparameter with the degree of influence on the overall effect.The multivariate Gaussian distribution is indicated as $h(\mathcal{N}(\mu,\sigma^2))$,which in turn can be formally expressed as:

\begin{equation}
    h(\mathcal{N}(\mu,\Sigma))=\frac{1}{2}log(det(2\pi e\Sigma))
    \label{eq:equation16}
\end{equation}

where $\Sigma$ indicates the covariance matrix, and since a diagonal array is used, the diagonal vector of $\Sigma$ is $\sigma^2$,and the above Eq.$\ref{eq:equation16}$ can be transformed into Eq.\ref{eq:equation17}:

\begin{equation}
    \begin{aligned}
        h(\mathcal{N}(\mu,\Sigma))  &= \frac{1}{2}log(det(2\pi e\Sigma))  \\ & = \frac{2}{d}(log(2\pi) + 1)+\displaystyle\sum_{i=1}^{d}log\sigma_i       
    \end{aligned}
    \label{eq:equation17}
\end{equation}

where d refers to the dimension of the feature.
For the final of the fake news detector, we use the Softmax function to convert the logistic values into category probabilities.Let:$\hat{P}=[\hat{P}^0_n,\hat{P}^0_n]$ be the final output probability vector and $Y_n$ be the ground-truth labeled value for $n-th$ news samples, where $\hat{P}^0_n$ and $\hat{P}^1_n$ denote the $n-th$ sample as the true and false predicted probabilities.The MDF framework will minimize the cross-entropy loss defined in Eq.\ref{eq:equation13} to end-to-end train the model to better detect fake news.Employing a penalty term with the standard normal distribution has been applied in other papers\cite{WeiHZH23}, and we compare in detail its performance with that of MDF using the entropy of the Gaussian distribution alone as the optimization function in subsection 4.6.3.

%-------------------------------------------------------------------------------- 4
\section{EXPERIMENTS}
\label{sec:EXPERIMENTS}
In this section, we first present the public dataset used in the experiments, as well as the basic setup of the experiments, while listing the common baseline models used for fake news detection.In addition, we experimentally evaluate the performance of the proposed dynamic fusion framework for fake news detection.The superior performance of the MDF framework is well demonstrated in relevant ablation experiments.Finally, we further explore the effective selection of each module and loss function by means of qualitative and quantitative analyses, and analyze the impact of hyperparameters on the model.

%--------------------------------------------------------------------------------4.1
\subsection{Datasets}
\label{sec:datasets}
In order to demonstrate the effectiveness of the proposed model, we experimented with our model using two public datasets, Twitter and Weibo.

\textbf{Twitter dataset}.This dataset was collected and published by the MediaEval Benchmarking Initiative (Boididou,Papadopoulos et al.\cite{boididou2014challenges}) to evaluate the performance of multi-modal models.Specifically, the dataset contains a large amount of social tweet content, which consists of text and images, and its training set consists of 6000 rumor posts and 5000 real posts.The test set contains different types of breaking news with up to 2000 posts.Posts containing only images or tweets are forcibly removed and are not allowed to participate in the testing process of the model.

\textbf{Weibo dataset}.This Chinese dataset has been widely used in many recent studies to verify the superior performance of their fake news detector.This dataset was first collected and publicly released by Jin et al.\cite{withRNN}, and contains only Chinese post and comment information.Sina Weibo actively encourages users to report any suspicious accounts and malicious speculations and comments on current events.As a result, a non-profit committee of reputable users distinguishes numerous news contents into real and fake news by manually verifying the cases.Many recent studies on detection systems were used as authoritative data sources and were further categorized and flagged by users from 2014 to 2016.Low-quality images and tweet posts are kept in our work to further demonstrate the robust performance of our classification model.The training set constitutes 70\% of our entire dataset and the rest is divided into validation and test sets in a 1:2 ratio.

Table\ref{tab:datasets} shows the statistical information of these two publicly available datasets.
% Table 1

\begin{table}
	\belowrulesep=0pt
	\aboverulesep=0pt
	\centering
	
	\begin{tabular}{cccc}
		\toprule
		Dataset	 & 	Label	 & 	Number	 & 	Total	 \\ 
		\midrule
		\multirow{2}{*}{Twitter}  & 	fake	 & 	7021	 & 	\multirow{2}{*}{12995}	 \\ 
		 & 	real	 & 	5974	 & 		 \\
		 \hline
		 \multirow{2}{*}{Weibo}	 & 	fake	 & 	4749	 & 	\multirow{2}{*}{9528}	 \\ 
		 & 	real	 & 	4779	 & 		 \\ 
		\bottomrule
	\end{tabular}
        \caption{Statics of two real-world multi-modal news datasets.}
        \label{tab:datasets}
\end{table}

%--------------------------------------------------------------------------------4.2
\subsection{Experiment Setups}
\label{sec:4.2}
\textbf{Parameter settings}.Adopting the common approach used in many previous studies\cite{WeiHZH23}\cite{SAFE}\cite{MVAE},etc.We initialize the text embeddings using bert-base-chinese and bert-base-uncased sentence-level and word-level annotations for the Weibo and Twitter datasets respectively,with the obtained feature dimensions of dt equals to 768.In order to avoid catastrophic forgetting of previous generalized knowledge\cite{french1993catastrophic}, the parameters of the BERT model\cite{devlin2018bert} were frozen for our model training.For the image data, we first standardized the image size to $224\times 224 \times 3$ and fed it into the pre-trained ResNet-50 network\cite{he2016deep}, after which it extracted features with a dimension of $d^v$ equal to 2048.In the UEM module, which is used to model intra-modal uncertainty, a linear network was employed to obtain the query, key, and value vectors within each modality.Prior to this, we used the idea of layer normalization.To prevent the risk of overfitting, dropout=0.1 was considered for the attention linear layer, and each linear layer was followed by a nonlinear transformation using the GELU activation function, employing 12 heads.Experimentally, it was shown that for the Twitter dataset, the threshold $\gamma$ in Algorithm \ref{alg:alg1} was set to 0.4 has excellent performance capability.Similarly, for the Weibo dataset, we chose to set the threshold $\gamma$ to 0.35.The reparameterized sample size was set to 5.The dropout of the graph attention layer was set to 0.4,while the ELU activation function was employed.After we embedded the fake news detector into the GAT structure,it consisted of 2 linear layers and employed a dropout value of 0.4 and ReLU activation function for nonlinear mapping.The model was trained on a batch size of 128 and 40 epochs with a learning rate initialized to 1e-4. 

\textbf{Evaluation Metrics}.We follow previous work\cite{withRNN} and use the F1 score as the final assessment.Specifically, the F1 score is calculated using the following formula:

\begin{equation}
        F1\,=\,\frac{2\times P\times R}{P+R}
    \label{eq:equation18}
\end{equation}

Where $P\!=\!\frac{TP}{TP+FP}$,$R\!=\!\frac{TP}{TP+FN}$.The accuracy is indicated as:
\begin{equation}
        ACC. \,=\, \frac{TP+TN}{TP+TN+FP+FN}
    \label{eq:equation19}
\end{equation}
Where $P$ stands for precision, $R$ stands for recall, $T\!P$ (True Positive) denotes the total number of posts predicted to be in the positive category, $F\!P$ (False Positive) refers to the total number of samples in the negative category that were predicted to have a positive label, $F\!N$ (False Negative) denotes the roundup of samples from the positive category that were predicted to be in the negative category, and $T\!N$ (True Negative) denotes the total number of samples that were predicted to be negative.We used the Adam optimizer\cite{kingma2014adam} to optimize our model parameters to make optimal decisions.
%--------------------------------------------------------------------------------4.3

\subsection{Baselines}
Most of the many previous research works are centered around unimodal fake news detection and multimodal fake news detection, so in order to better evaluate the superiority of our proposed model, we will also compare it with the baseline model that employs a single textual modality, i.e.,we eliminate the dynamic decision-making part and directly input the noisy single-textual posts into the UEM module, and splicing a fully connectivity layer is used to output the predicted values.
% Table 2

\begin{table*}
	\belowrulesep=0pt
	\aboverulesep=0pt
	\centering
	\caption{Performance comparison of MDF proposed by us with other Baselines on both Twitter and Weibo datasets.}
	\begin{tabular}{|c|c|c|c|c|c|c|c|c|}
		\toprule
		\multirow{2}{*}{Dataset} & \multirow{2}{*}{Method} & \multirow{2}{*}{ACC} & \multicolumn{3}{c|}{Fake News} & \multicolumn{3}{c|}{Real News} \\
		\cline{4-6}\cline{7-9}
		& & & P & R & F1 & P & R & F1 \\
		\midrule
		\multirow{12}{*}{Twitter} & 		SVM-TS\cite{ma2015detect}	 & 	0.529	 & 	0.488	 & 	0.497	 & 	0.496	 & 	0.565	 & 	0.556	 & 	0.561	 \\ 
		& 		GRU\cite{ma2016detecting}	 & 	0.634	 & 	0.581	 & 	0.812	 & 	0.677	 & 	0.758	 & 	0.502	 & 	0.604	 \\ 
		& 		CNN\cite{yu2017convolutional}	 & 	0.549	 & 	0.508	 & 	0.597	 & 	0.549	 & 	0.598	 & 	0.509	 & 	0.550	 \\ 
		& 		BERT\cite{devlin2018bert}	 & 	0.706	 & 	0.648	 & 	0.540	 & 	0.589	 & 	0.715	 & 	0.636	 & 	0.673	 \\ 
            \cline{2-9}
		& 		SAFE\cite{SAFE}	 & 	0.766	 & 	0.777	 & 	0.795	 & 	0.786	 & 	0.752	 & 	0.731	 & 	0.742	 \\ 
		& 		VQA\cite{VQA}	 & 	0.631	 & 	0.765	 & 	0.509	 & 	0.611	 & 	0.550	 & 	0.794	 & 	0.650	 \\ 
		& 		MVAE\cite{MVAE}	 & 	0.745	 & 	0.801	 & 	0.719	 & 	0.758	 & 	0.689	 & 	0.777	 & 	0.730	 \\ 
		& 		EANN\cite{Eann}	 & 	0.648	 & 	0.810	 & 	0.498	 & 	0.617	 & 	0.584	 & 	0.759	 & 	0.660	 \\ 
		& 		SpotFake\cite{SpotFake}	 & 	0.892	 & 	0.902	 & 	0.964	 & 	0.932	 & 	0.847	 & 	0.656	 & 	0.739	 \\ 
		& 		BDANN\cite{zhang2020bdann}	 & 	0.821	 & 	0.790	 & 	0.610	 & 	0.690	 & 	0.830	 & 	0.920	 & 	0.870	 \\ 
		& 		MCNN\cite{Byconsistency}	 & 	0.823	 & 	0.858	 & 	0.801	 & 	0.828	 & 	0.787	 & 	0.848	 & 	0.816	 \\ 
		& 		\textbf{MDF}	 & 	\textbf{0.947}	 & 	\textbf{0.876}	 & 	\textbf{0.962}	 & 	\textbf{0.917}	 & 	\textbf{0.983}	 & 	\textbf{0.941}	 & 	\textbf{0.961}	 \\ 
		\hline
		\multirow{9}{*}{Weibo} & 	SVM-TS\cite{ma2015detect}	 & 	0.640	 & 	0.741	 & 	0.573	 & 	0.646	 & 	0.651	 & 	0.798	 & 	0.711	 \\ 
		& 	GRU\cite{ma2016detecting}	 & 	0.702	 & 	0.671	 & 	0.794	 & 	0.727	 & 	0.747	 & 	0.609	 & 	0.671	 \\ 
		& 	CNN\cite{yu2017convolutional}	 & 	0.740	 & 	0.736	 & 	0.756	 & 	0.744	 & 	0.747	 & 	0.723	 & 	0.735	 \\ 
		& 	BERT\cite{devlin2018bert}	 & 	0.804	 & 	0.800	 & 	0.860	 & 	0.830	 & 	0.840	 & 	0.760	 & 	0.800	 \\ 
            \cline{2-9}
		& 	VQA\cite{VQA}	 & 	0.736	 & 	0.797	 & 	0.634	 & 	0.706	 & 	0.695	 & 	0.838	 & 	0.760	 \\ 
		& 	MVAE\cite{MVAE}	 & 	0.824	 & 	0.854	 & 	0.769	 & 	0.809	 & 	0.802	 & 	0.875	 & 	0.837	 \\ 
		& 	EANN\cite{Eann}	 & 	0.794	 & 	0.790	 & 	0.820	 & 	0.800	 & 	0.800	 & 	0.770	 & 	0.780	 \\ 
		& 	BDANN\cite{zhang2020bdann}	 & 	0.814	 & 	0.800	 & 	0.860	 & 	0.830	 & 	0.840	 & 	0.760	 & 	0.800	 \\ 
		& 	\textbf{MDF}	 & 	\textbf{0.819}	 & 	\textbf{0.721}	 & 	\textbf{0.652}	 & 	\textbf{0.685}	 & 	\textbf{0.855}	 & 	\textbf{0.891}	 & 	\textbf{0.873}	 \\ 
		\bottomrule
	\end{tabular}
	\label{tabl2}
\end{table*}

\begin{enumerate}[(a)]
  \item Fake news detection model based on single text modality. SVM-TS\cite{ma2015detect} employs a machine learning approach to detect fake news using heuristic rules and linear multi-layer perceptron classifiers. In order to learn the temporal feature information for fake news detection, a recurrent neural neural network implementation is deployed in GRU\cite{ma2016detecting} to model it.CNN\cite{yu2017convolutional} is employed to learn the feature representation of fake news detection by framing related posts into fixed-length sequences was introduced in \cite{yu2017convolutional}. Additionally, we constructed a model based on the pre-trained BERT\cite{devlin2018bert} with a fully connected layer behind the BERT for detecting fake news with plain text input.
  \item Fake news detection model based on multimodal features. VQA\cite{VQA} aims to answer questions about the given images. The original VQA model is designed for multi-class classification tasks.The SAFE\cite{SAFE} proposes a similarity-aware cross-modal fusion function for multi-modal FND..The MVAE model\cite{MVAE} extracts modal information from text and images and uses a multimodal variant encoder for simple fusion.\cite{Eann} constructs an Event Adversarial Neural Network (EANN) that uses an event learner to learn the feature representations of text and images in an article, but adding additional auxiliary features will increase the cost of detection. Intuitively, the SpotFake model\cite{SpotFake} can use large-scale textual and visual pre-trained models to extract features for each modality and concatenate them into a multimodal representation of the news, which is then placed into a binary classifier for detecting fake news.MCNN\cite{Byconsistency} considers the consistency of different modalities, and captures the global characteristics of social media information.BDANN\cite{zhang2020bdann} proposed a BERT-based domain adaptive neural network for detecting multimodal fake news.
\end{enumerate}
%--------------------------------------------------------------------------------4.4
\subsection{Main Results}
This subsection compares the performance of the dynamic fusion framework called MDF proposed by us with the existing baselines, while we give specific analytical results.
The experimental results in Table\ref{tabl2} fully illustrate the superiority of our proposed framework on two publicly available datasets. Specifically, we can derive the following results:

\begin{enumerate}[(a)]
  \item Among the unimodal methods, especially on textual modalities, the pre-trained language models(PLMs) represented by BERT shows its irreplaceable capturing ability. On the Twitter dataset, it outperforms the traditional Support Vector Machine (SVM) by almost 9.3\%. This reflects the effectiveness of employing textual unimodality for fake news detection, as visual fakery is usually not easily captured.
  \item Multimodal fake news detection methods (e.g., MVAE, EANN, etc.) usually have more sensitive insights than unimodal text detection methods because they usually take into account the inconsistent semantic information of both modalities, text and image, in their models. Thus, this is a good proof of the effectiveness of employing multimodal features for the task of fake news detection.
\end{enumerate}

On two public datasets, Twitter and Weibo, our proposed dynamic fusion framework named MDF consistently outperforms the latest baseline methods. A 10.2\% improvement in accuracy over the optimal baseline model is realized.Correspondingly, the F1 Score rises by 15.6\%. In this paper, the data uncertainty problem caused by the noise in the multimodal posts circulating in social networks is elaborated, while the uncertainty modeling mechanism is introduced in our model, which is able to capture the unimodal features in the complex heterogeneous multimodal post data, and in addition, the heterogeneous graph attention network shifts the goal to modeling the inter-modal uncertainty, and the mapped features have stronger robustness, so the performance is effectively improved.

%--------------------------------------------------------------------------------4.5
\subsection{Ablation Studies}
To verify the effectiveness of our proposed fusion strategy, we conducted separate ablation analyzes to evaluate the effectiveness of each component in improving the performance of the MDF framework. For each experiment, we removed a different component UEM or DFN and started training the model again. A variant of MDF was implemented as follows:

\begin{enumerate}
  \item MDF w/o UEM-Text. the uncertainty module used to model unimodal text was removed. We feed the UEM module features with captured visual uncertainty information and features extracted after BERT\cite{devlin2018bert} into the dynamic fusion module, and we do not remove Gaussian modeling for text features in order to better represent the modeling capabilities of DFN.
  \item MDF w/o UEM-Image.The uncertainty module for modeling unimodal images was removed.We will use the UEM module features with captured textual uncertainty information and extracted features from the pre-trained ResNet-50 network to feed into the dynamic fusion module.Similarly, Gaussian modeling for visual features was not removed.
  \item MDF w/o UEM. the uncertainty module used to model unimodal text and images is eliminated. We assume that the two unimodal modalities, text and image, have the same confidence score and feed directly into the dynamic fusion module using BERT\cite{devlin2018bert} and a simple splicing of text and image features extracted by the pre-trained ResNet-50 network.
  \item MDF w/o DFN. the dynamic fusion module is removed. The decision-making capability of the model relies only on modeling robust intra-modal uncertainty features extracted from unimodal text and unimodal images by simple splicing operations.
\end{enumerate}

\begin{figure*}[!t]
    \centering
    \includegraphics[width=1\columnwidth]{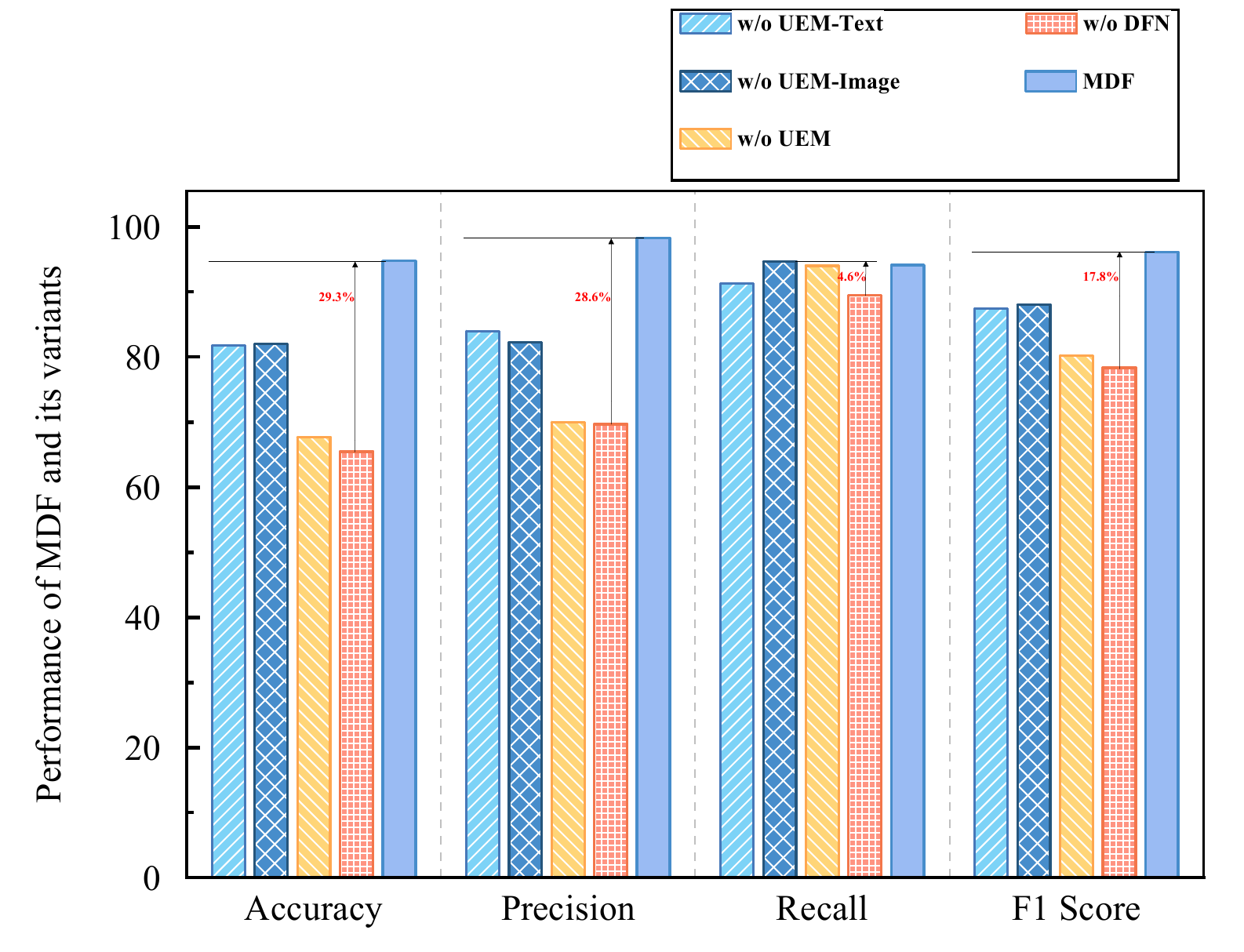}
    \caption{Visualize the resultant comparative performance of MDF with its variants.}
    \label{fig:ablation}
\end{figure*}

We visualize the results of our experiments on the Twitter dataset as shown in Fig.\ref{fig:ablation}.The overall performance of the uncertainty modeling modules excluding either unimodal text or unimodal image, i.e., w/o UEM-Text and w/o UEM-Image, are both degraded. Specifically, excluding the uncertainty modeling module for unimodal text, accuracy shows a significant dip after w/o UEM-Text. This exemplifies the extent to which unimodal text affects the accuracy of the fake news detection task, and further illustrates the importance of uncertainty modeling. Similarly, when the uncertainty module used to model unimodal images was removed, the overall model performance also showed a dip. However, the extent of the effect of the above two compared to that after the direct deletion of the UEM module, i.e., w/o UEM, is smaller. When the UEM module is directly deleted, the overall performance shows a sharp decline. The w/o DFN that just relies on simple splicing instead of dynamic fusion strategy shows different degrees of degradation in both Accuracy as well as F1 score.

It is clear to observe that the effect of excluding the DFN of our proposed dynamic fusion architecture realizes poor results, with the worst Accuracy directly decreasing by up to \textbf{29\%}, which further illustrates the superior performance of our proposed dynamic fusion architecture. We attribute this to the superiority of the dynamic fusion framework, which is capable of adjusting the final decision-making strategy based on the dynamic weights held by each modality, in particular, the use of normalized coefficient of variation as the fusion weights. Secondly, the performance of the w/o UEM variant used to indicate deletion of uncertainty modeling is also significantly degraded, which is a good indication of the adaptability of uncertainty modeling to our dynamic fusion framework.

%--------------------------------------------------------------------------------4.6
\subsection{Qualitative Analysis}
In this subsection, in order to prove the validity of our model, we conducted a lot of experiments to validate it, the first one is to combine the traditional cross-attention mechanism with our DFN module to form an MDF framework using the cross-attention mechanism, but poorly it does not work as well as we would like. This experiment is mainly used to show that our proposed UEM module is a perfect fit with the DFN module, and the role of uncertainty modeling in MDF cannot be replaced by the traditional simple attention mechanism. In addition, we converted the DFN module to many previously used methods, but the results still did not achieve the excellent performance using our proposed dynamic fusion strategy.
%--------------------------------------------------------------------------------4.6.1
\subsubsection{MDF using cross-attention mechanism vs. using graph-attention mechanism}
\begin{figure*}[!t]
\centering
\subfloat[MDF with cross-attention.]{\label{fig_6.1}
		\includegraphics[width=1\columnwidth]{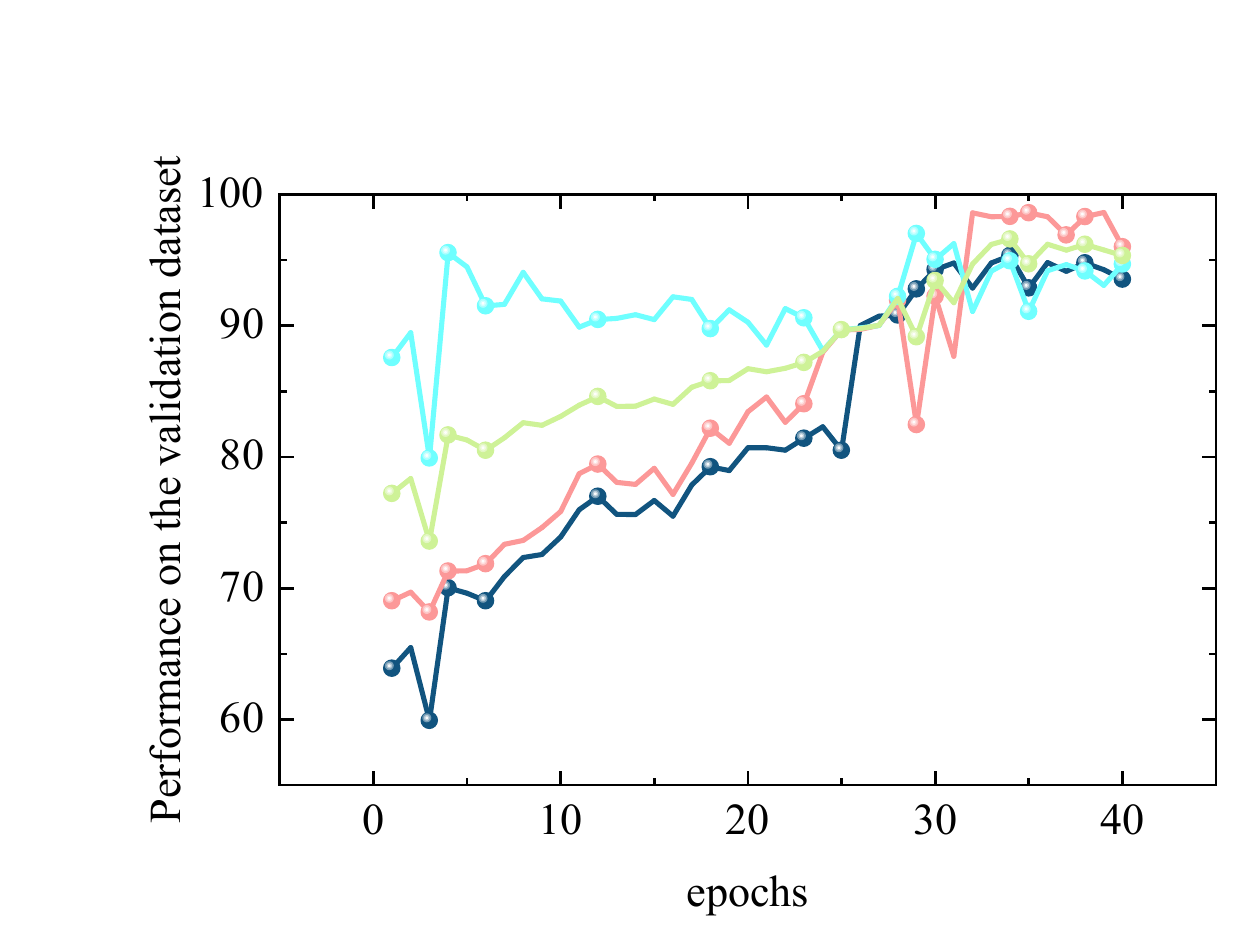}}
\subfloat[MDF with graph attention.]{\label{fig_6.2}
		\includegraphics[width=1\columnwidth]{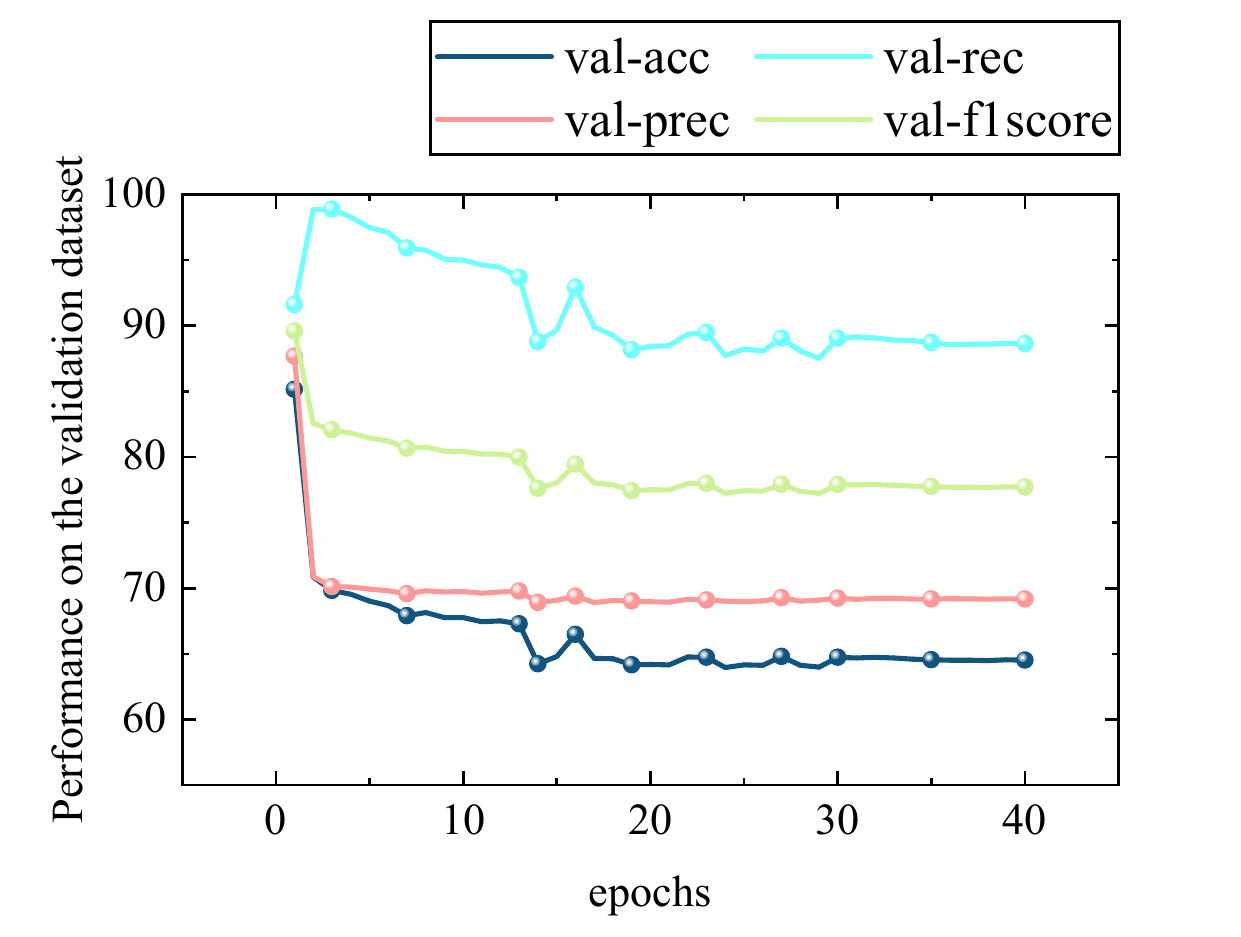}}
\caption{Where val-acc refers to the Accuracy of each model on the test set, val-prec refers to the Accuracy of each model on the test dataset, val-rec refers to the Recall of each model on the test dataset, and val-f1score refers to the F1 score of each model on the test dataset. note that the above experimental results all are experimental data on Twitter.}
\label{fig_6}
\end{figure*}

We also continue the approach of employing cross-attention for using in the MDF framework as proposed by numerous previous researchers, but obtained very poor experimental results, as shown in Fig.\ref{fig_6}\subref{fig_6.1}.

As can be seen in Fig. \ref{fig_6}\subref{fig_6.1}, the value of F1 score constantly shows a stepwise decrease during the model testing phase, while accuracy and precision show different degrees of decrease. This indicates that the use of the traditional cross-attention mechanism does not fit well with our proposed dynamic fusion framework. On the contrary, using the graph attention mechanism can substantially improve the effectiveness of the model, the effect of which can be seen in \ref{fig_6}\subref{fig_6.2}. Although there is an unanticipated decrease in efficiency in a small region, the overall trend of the model is positive, and the F1 score in its best performance result is improved by \textbf{7\%} over the effect of using cross-attention.Table\ref{tabl3} documents the best- and worst-result performance capabilities of the MDF framework with cross-attention versus the MDF framework with the graph attention mechanism on the Twitter dataset. We attribute this to the fact that the graph attention mechanism can better capture robust feature representations obtained using unimodal uncertainty modeling, whereas the cross-attention-only mechanism does not represent the heterogeneous properties of features well, while not being able to model inter-modal uncertainty.

\begin{table*}
	\belowrulesep=0pt
	\aboverulesep=0pt
	\centering
	\caption{Comparison of optimal results on Twitter dataset between MDF framework using cross-attention and using graph attention mechanism.}
	\begin{tabular}{cccccc}
		\toprule
		Model	 & 		 & 	Accuracy	 & 	Micro Precision	 & 	Micro Recall	 & 	Micro F1-Score	 \\ 
		\midrule
		\multirow{2}{*}{MDF-GAT}	 & 	max	 & 	94.791	 & 	98.313	 & 	94.159	 & 	96.191	 \\ 
		& 	min	 & 	59.924	 & 	68.186	 & 	79.908	 & 	73.583	 \\
		\hline
		\multirow{2}{*}{MDF-Att}	 & 	max	 & 	85.147	 & 	87.668	 & 	91.596	 & 	89.589	 \\ 
		& 	min	 & 	63.986	 & 	69.100	 & 	87.515	 & 	77.224	 \\ 
		\bottomrule
	\end{tabular}
	\label{tabl3}
\end{table*}

%--------------------------------------------------------------------------------4.6.2
\subsubsection{Effectiveness of the Dynamic Fusion Framework}
In order to illustrate the uncertainty modeling ability of the proposed dynamic fusion framework for robust representation features of single modalities, we eliminate the dynamic fusion module DFN and replace it with the following two methods:

$\bullet$ \textbf{MDF-Concat} implies a simple connection that performs multimodal features.

$\bullet$ \textbf{MDF-FC} uses a fully connected layer to fuse multimodal features.

$\bullet$ \textbf{MDF-Att} fuses multimodal features and attention mechanisms.

$\bullet$ \textbf{MDF-CoA} fuses multimodal feature information and common attention mechanisms.

\begin{figure}[!t]
    \centering
    \includegraphics[width=1\columnwidth]{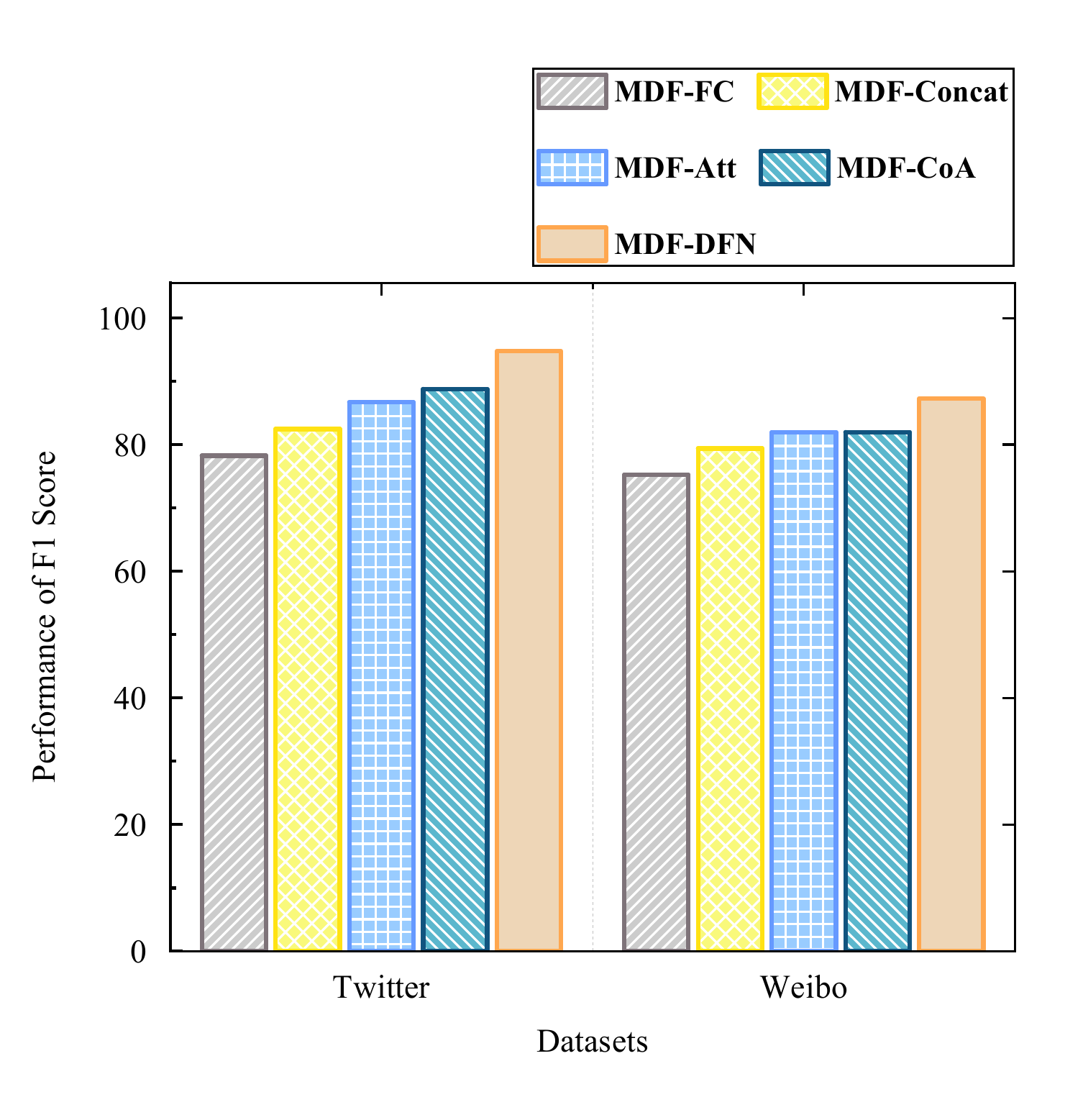}
    \caption{Performance of the MDF framework with different fusion strategies on two public datasets.}
    \label{fig:dfn}
\end{figure}

Fig.\ref{fig:dfn} shows the performance of MDF with different fusion strategies. From the experimental results, it can be seen that the method using the DFN module as the fusion strategy for MDF combined with uncertainty modeling outperforms other fusion modules on both datasets, which demonstrates the superior performance of our model.The performance of MDF under the MDF-FC and MDF-Concat mechanisms is slightly inferior to that of MDF-Att and MDF-CoA using the Attention mechanism , which also fully illustrates the strong ability of the attention mechanism in capturing complementary feature information. Meanwhile, the fully connected layers and simple splicing operations are also limited by the fact that they ignore the complex logical relationships between text and visual modalities.

%--------------------------------------------------------------------------------4.6.3

\subsubsection{Comparison of Loss Functions}
The Kullback-Leibler scatter (K-L scatter) is a measure of the difference between two probability distributions. In uncertainty modeling, the KL scatter is commonly used to compare the level of differentiation between a modeled probability distribution and a standard normal distribution, which is attributed to the fact that data uncertainty tends to be modeled in the form of a normal distribution. Formally, the penalty term using K-L scatter can be defined as:

{
\begin{equation}
        \begin{aligned}
            \mathcal{L}_{kl} &= KL(\mathcal{N}(\mathcal{z};\mu,\sigma^2)||\mathcal{N}(\epsilon;0,I)) \\ &= -\frac{1}{2}(1+log(\sigma^2)-(\mu)^2-(\sigma)^2)
            \label{eq:equation20}   
        \end{aligned}
\end{equation}
}

In the formula,$KL(\cdot||\cdot)$ indicates the KLD between two probability distributions.

\begin{figure}[!t]
    \centering
    \includegraphics[width=1\columnwidth]{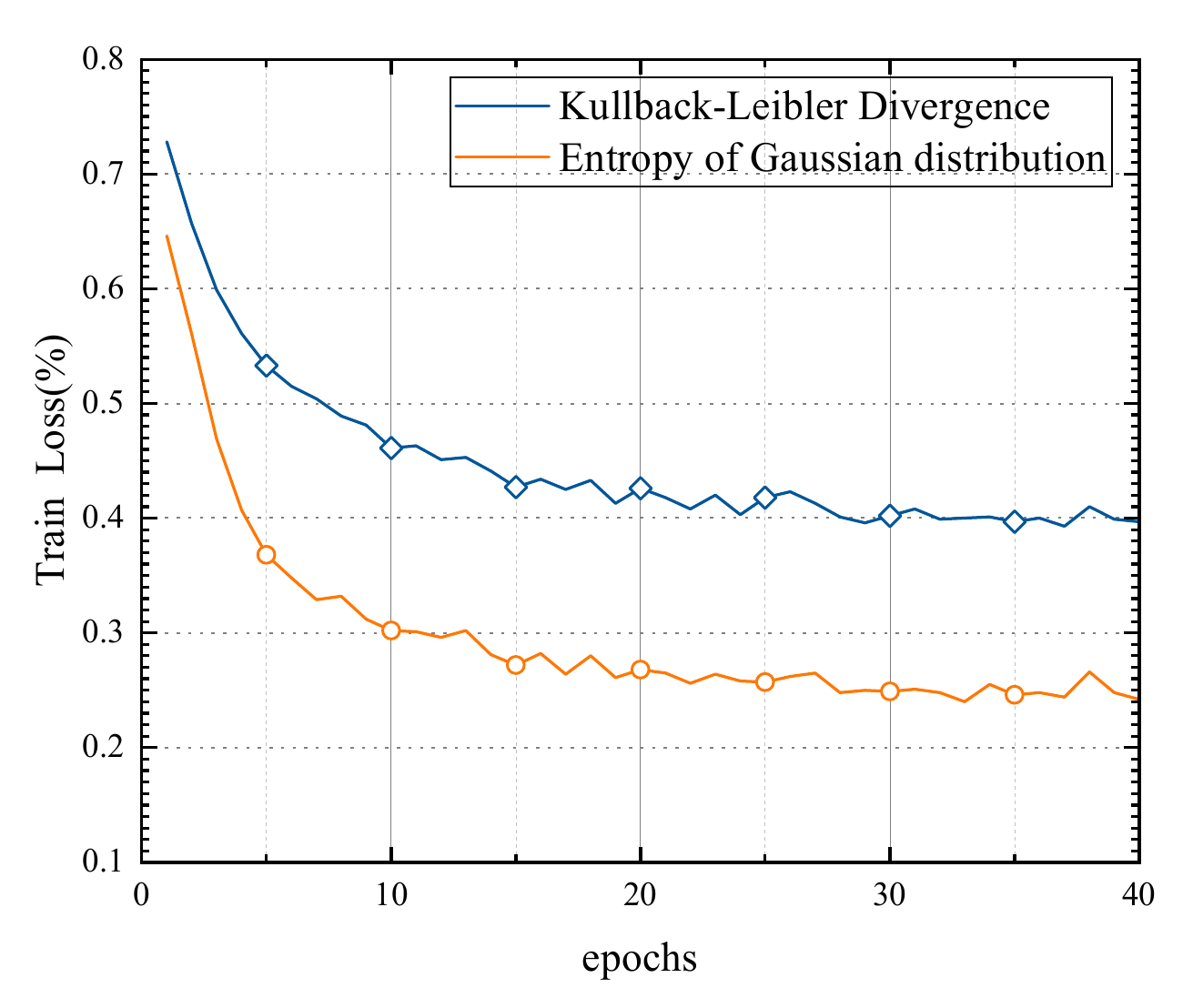}
    \caption{Performance of the MDF framework with different fusion strategies on two public datasets.}
    \label{fig:lossfunc}
\end{figure}

When the value of K-L scatter is small, it indicates that the degree of differentiation between the model distribution and the standard normal distribution is small, and the model fits the distribution of the real data well. Therefore, the model's ability to model uncertainty can be continuously improved by optimizing the model to reduce the K-L scatter when modeling uncertainty. We also adopt the Kullback-Leibler scatter as the optimization function of the MDF framework, so we redefine the optimization algorithm of the MDF as shown in Eq.\ref{eq:equation21}. Its comparison with entropy using Gaussian distribution is shown in Fig.\ref{fig:lossfunc}, which is a metric that describes the uncertainty of a random variable of a certain distribution.

{
\begin{equation}
        \begin{aligned}
            \mathcal{L}_{pred}=\mathcal{L}(\theta)+\alpha\mathcal{L}_{kl}
            \label{eq:equation21}   
        \end{aligned}
\end{equation}
}
In the formula,$\mathcal{L}(\theta)$ is still the cross-entropy loss function for the fake news detection task defined by Eq.\ref{eq:equation13}.And $\alpha$ is the penalty term coefficient.

As can be seen from Fig.\ref{fig:lossfunc}, the training loss of the MDF framework using K-L scattering is larger than that of the MDF framework using the entropy of the Gaussian distribution as the penalty term, especially in the later stages of model training, where it fails to converge to a reasonable interval over time. In addition, comparing the convergence speeds to reach the optimal state, the convergence speed of the framework employing K-L dispersion is relatively slow compared to the framework employing the entropy of the Gaussian distribution as the loss. Specifically, the fusion framework using the entropy of Gaussian distribution as the penalty term has gradually converged into the optimal interval in the 15th epochs, but this is relatively difficult for the model using K-L scatter. Meanwhile, in the late stage of model training, the MDF framework utilizing the entropy of the Gaussian distribution can well predict the fluctuation of the data in the unknown interval, but the framework using K-L scattering does not perceive it well. In summary, both in terms of convergence speed and performance capability, using the entropy of Gaussian distribution as the final penalty term is much better than using Kullback-Leibler divergence (K-L divergence).We attribute this to the fact that the entropy of a Gaussian distribution can encourage the model to generate a broad distribution rather than one similar to the standard normal distribution. This is beneficial in Bayesian optimization because the model will be forced to quantify the uncertainty of the prediction.

\begin{figure*}[!t]
\centering
\subfloat[Twitter dataset.]{\label{fig_9.1}
		\includegraphics[width=1\columnwidth]{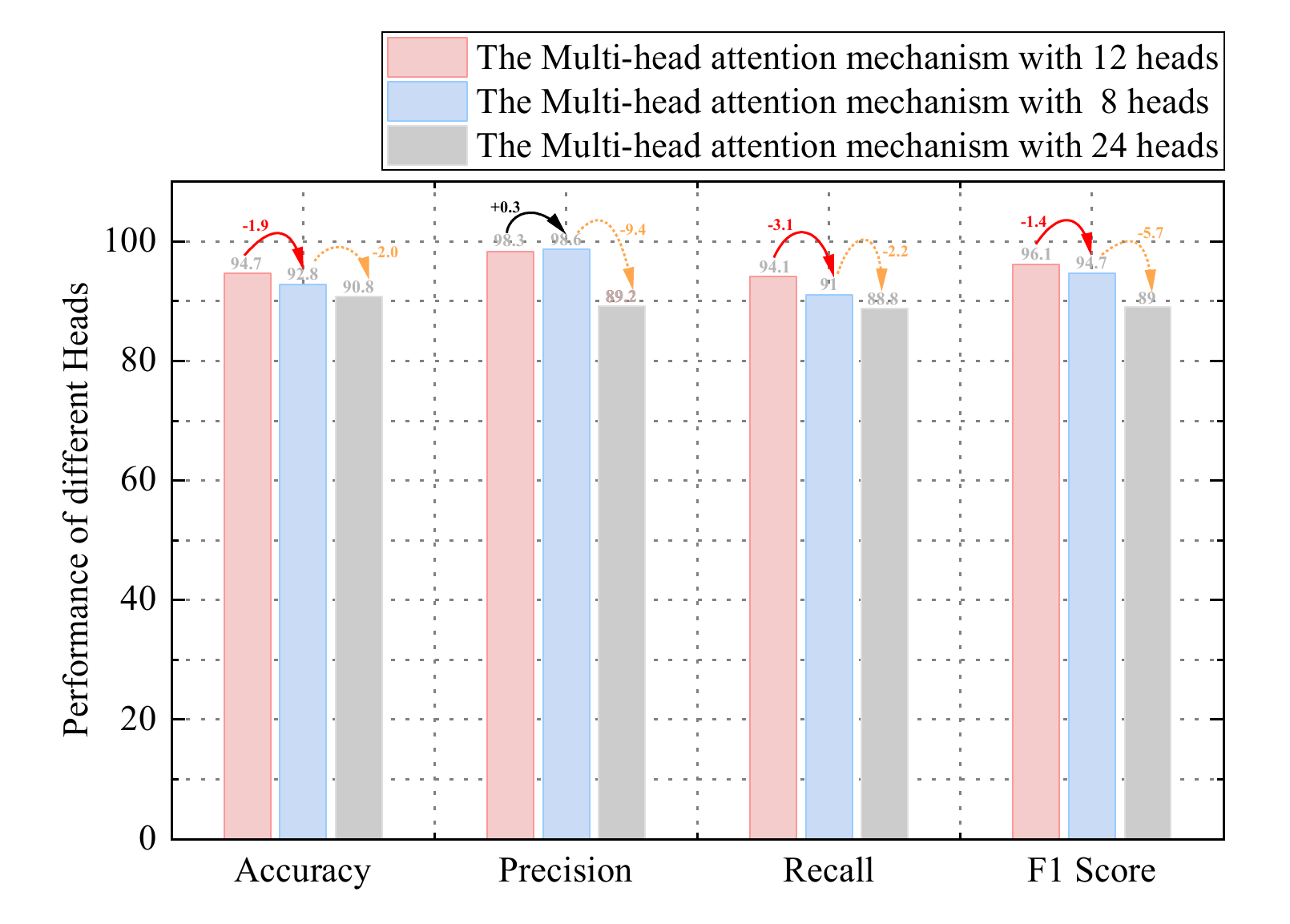}}
\subfloat[Weibo dataset.]{\label{fig_9.2}
		\includegraphics[width=1\columnwidth]{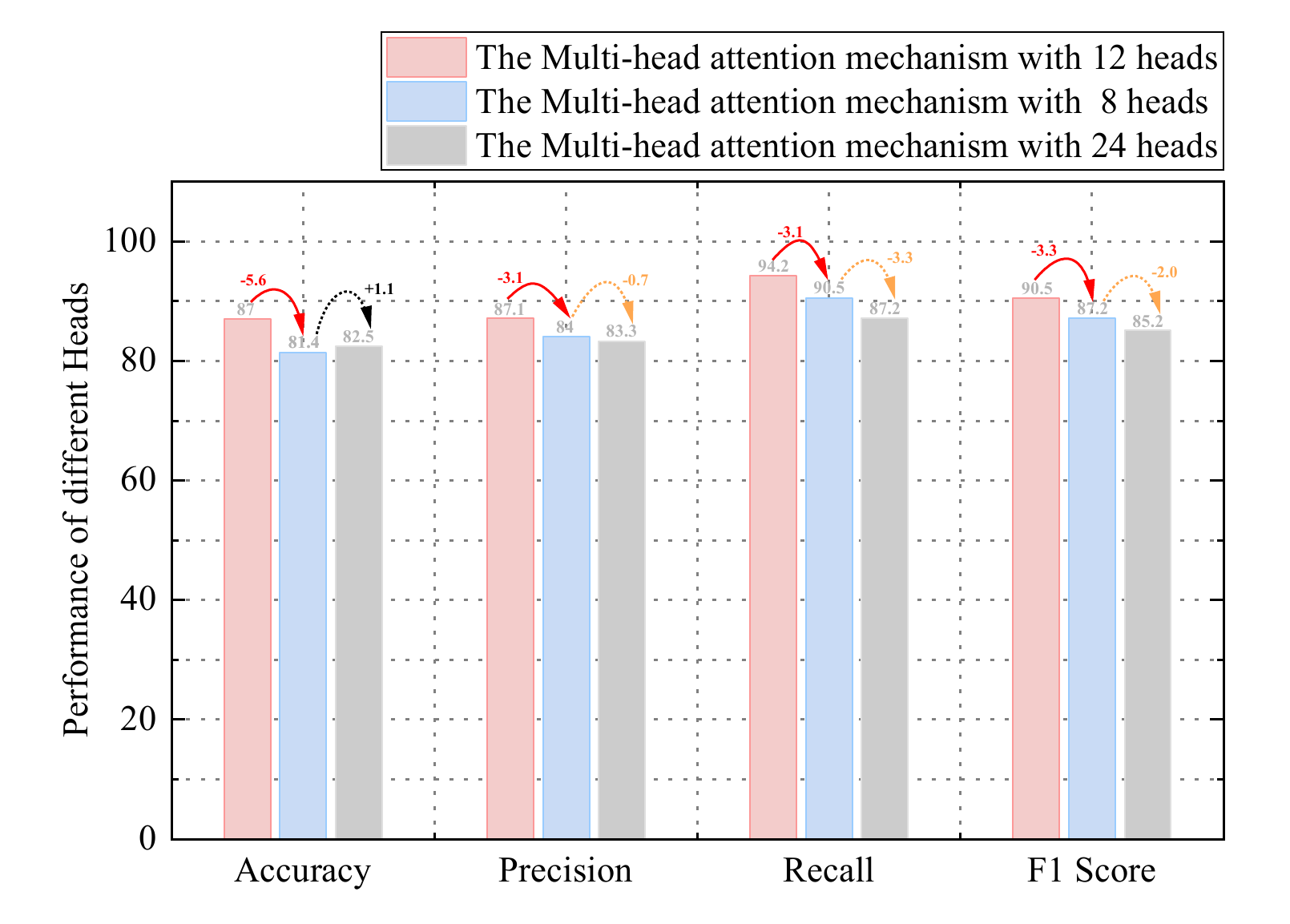}}
\caption{Comparison on Twitter and Weibo dataset using multi-head attention mechanism with 12 heads,8 heads and 24 heads.}
\label{fig:MultiHead}
\end{figure*}

%--------------------------------------------------------------------------------4.7

\subsection{Quantitative analysis}
\label{sec:4.7}
In this subsection, we analyze in detail the role of the relevant hyperparameters involved in the model and their impact on the final effect of the MDF framework.

%--------------------------------------------------------------------------------4.7.1

\subsubsection{Parameter analysis of the attention mechanism in the UEM module}
The UEM module is mainly responsible for uncertainty modeling, in which we mainly use the multi-head attention mechanism. In order to evaluate the impact of the number of heads in the multi-head attention mechanism on the overall performance, we conducted comparative experiments on the Twitter and Weibo datasets, and the results are shown in Fig.\ref{fig:MultiHead}. The results show that on both the Twitter and Weibo public datasets, excellent performance can be achieved when the number of heads in the multi-head attention mechanism in UEM is controlled to be 8. However, as the number of heads decreases, the overall performance results also show a small decrease.

In Fig.\ref{fig:MultiHead}\subref{fig_9.1}, employing a multi-head attention mechanism with 12 heads on the Twitter dataset is identified as a promising strategy for modeling uncertainty, although it has a performance degradation of 0.3 in terms of Precision compared to the multi-head attention mechanism with 8 heads.Meanwhile, the effect on the Weibo data showed in Fig.\ref{fig:MultiHead}\subref{fig_9.2}.It shows the superiority of employing the 12-head multi-head attention mechanism is superior to 8-head attention, both in terms of accuracy and F1 score. However, it is not the case that more heads of multi-head attention are more effective, and we conducted further experiments with a bad drop in the results after setting the number of attention heads to 24.

%--------------------------------------------------------------------------------4.7.2
\subsubsection{Parameter analysis in the DFN module}
\label{sec:Analysis Dfn Params}
As the most important part of the dynamic fusion framework, the DFN module, which adopts the graph attention mechanism and D-S evidence theory, has a direct impact on the final dynamic decision. And in order to better balance the contribution of the two modalities, we take the normalized coefficient of variation of each modality as an important measure, and compare it with the hyperparameter in the DFN, which also serves as the $\gamma$ of the confidence threshold. Thus, this hyperparameter also directly influences the final dynamic fusion strategy. We conducted several sets of experiments on the Twitter dataset to verify the impact of this parameter on the final classification results. The experimental results are shown in Table\ref{tabl4}

\begin{table*}
	\belowrulesep=0pt
	\aboverulesep=0pt
	\centering
	\caption{The impact of the threshold $\gamma$ used for dynamic fusion in the DFN module on the overall effect on two publicly available datasets, Twitter and Weibo.}
        \begin{tabular}{cccccc}
		\toprule
		Dataset	 & 	$\gamma$(Threshold)	 & 	Accuracy	 & 	Precision	 & 	Recall	 & 	F1-Score	 \\
		\midrule
		\multirow{3}{*}{Twitter}	 & 	0.25	 & 	0.76	 & 	0.79	 & 	0.90	 & 	0.84	 \\
		& 	0.50	 & 	0.94	 & 	0.98	 & 	0.94	 & 	0.96	 \\
		& 	0.75	 & 	0.92	 & 	0.98	 & 	0.91	 & 	0.94	 \\
		\hline
		\multirow{3}{*}{Weibo}	 & 	0.25	 & 	0.76	 & 	0.79	 & 	0.90	 & 	0.84	 \\
		& 	0.50	 & 	0.81	 & 	0.85	 & 	0.89	 & 	0.87	 \\
		& 	0.75	 & 	0.72	 & 	0.73	 & 	0.94	 & 	0.82	 \\
		\bottomrule
	\end{tabular}
	\label{tabl4}
\end{table*}

The DFN module plays a pivotal role in the MDF framework. As the key hyperparameter $\gamma$ used for dynamic fusion by the DFN module, it mainly assumes the role of selecting the final strategy. Relying on the strong adaptation capability of the heterogeneous graph structure, when a certain modality is filled with a large amount of noise and data uncertainty occurs, the DFN module decides whether or not to classify it with the unimodal uncertainty modeling capability through the threshold $\gamma$. As can be seen from Table\ref{tabl4},the value of $\gamma$ should be chosen to fit exactly a certain distribution satisfied by the dataset, and a large number of randomized experiments should be carried out for specific datasets to finalize the threshold.We conducted six sets of experiments with $\gamma$ values of 0.25, 0.5 and 0.75, transformed at 50\% and 30\%, respectively.The experimental results show that the effect of $\gamma$ values on the Weibo dataset is small relative to the Twitter dataset,with the maximum difference in F1 reaching \textbf{0.05\%}. Whereas, for the Twitter dataset, there is some difference in the selection of $\gamma$ values, with the best performance capability being \textbf{0.10\%} more than the worst.

%--------------------------------------------------------------------------------4.7.3

\subsubsection{Parameters in the Optimization Algorithm}
\begin{figure}[!t]
\centering
\subfloat[Twitter dataset.]{\label{fig_10.1}
		\includegraphics[width=0.5\columnwidth]{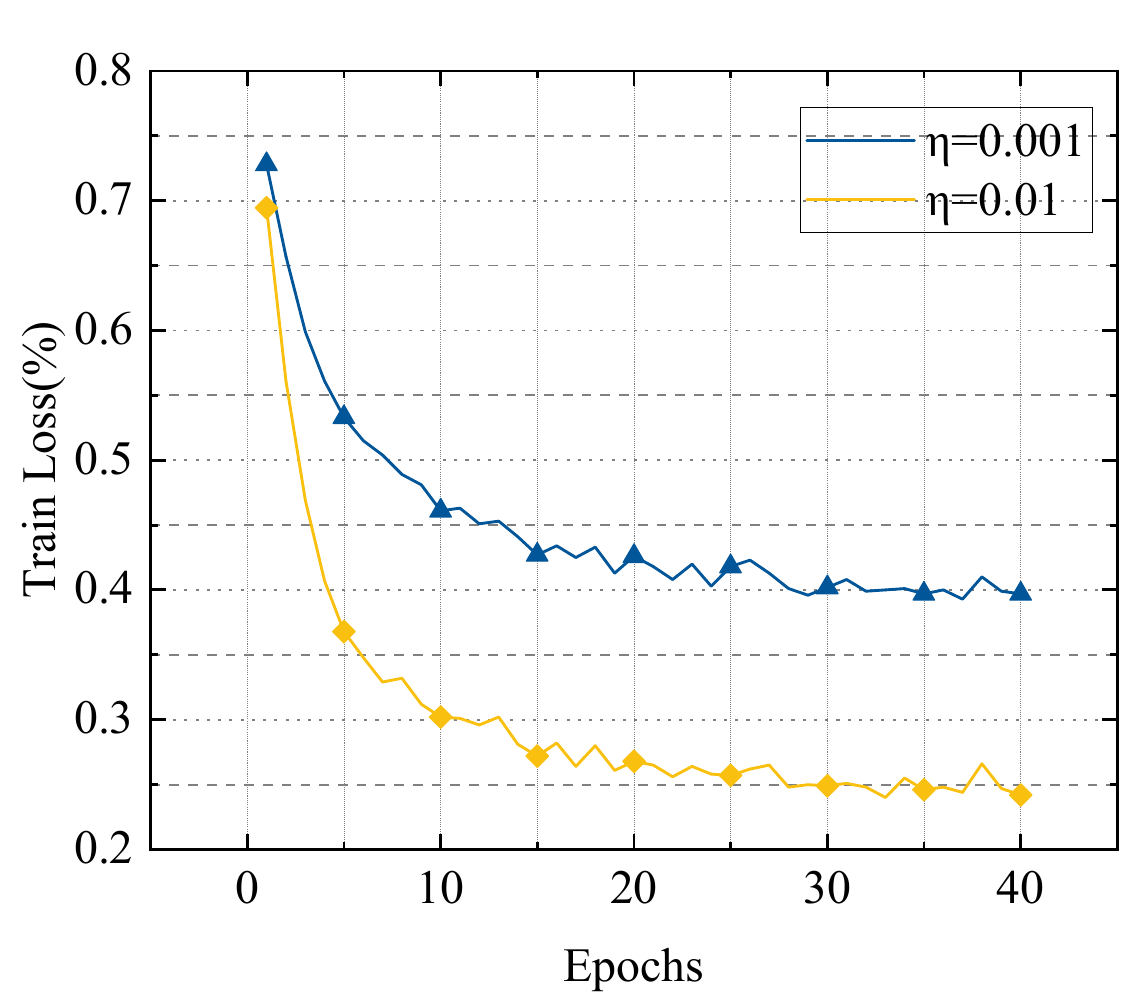}}
\subfloat[Weibo dataset.]{\label{fig_10.2}
		\includegraphics[width=0.5\columnwidth]{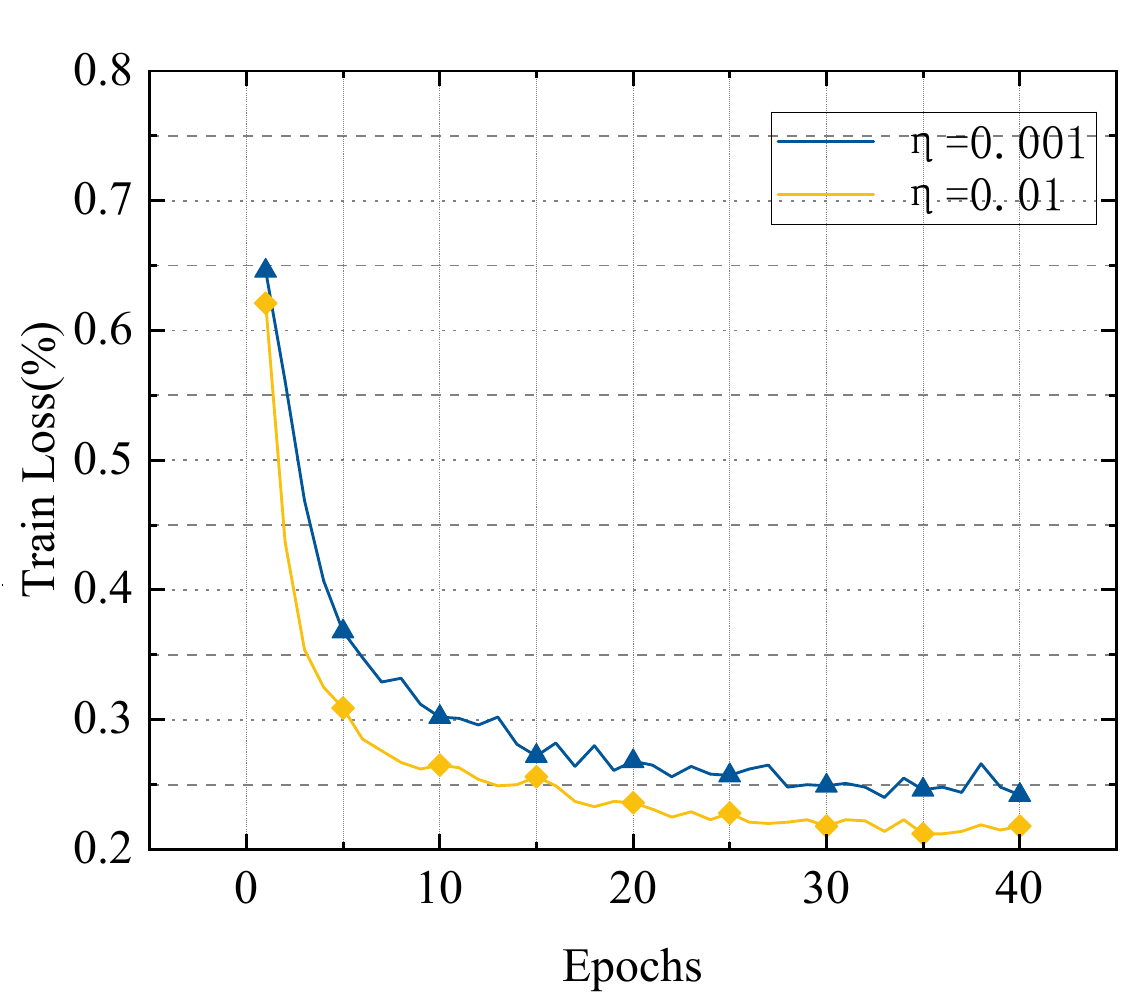}}
\caption{Performance of the MDF framework on the Twitter and Weibo dataset for hyperparameters $\eta$ taken as 0.001 and 0.01, respectively.}
\label{fig:OA}
\end{figure}

The cost function is responsible for evaluating the variance between the MDF framework and the real results, and hence its performance. Using only the cross-entropy loss function will lead to the collapse of the modeled variance and the degradation of the distribution representation from the original high-dimensional Gaussian distribution to a deterministic point embedding approach, which further leads to a reduction in the model's generalization ability for multimodal information. Therefore, we add the entropy of the Gaussian distribution as a penalty term to reduce the impact of modal uncertainty on the final result. The hyperparameter , as the upper bound value of the Gaussian distribution, directly affects the final decision-making effect, so we conducted several sets of experiments to select a reasonable value that is specifically applied to the two public datasets of Twitter and Weibo, and the results of the experiments are shown in Fig.\ref{fig:OA}.

As can be seen from Fig.\ref{fig:OA}, different $\eta$ values have some influence on the final results of both datasets, especially on the Twittet dataset, when $\eta$ takes the value of \textbf{0.01}, the MDF generalization ability is better than when $\eta=0.001$. We analyze the reason for this as a lower value of $\eta$ causes the penalty term containing a Gaussian distribution to eventually converge to zero or negative values. From Eq.\ref{eq:equation15}, this would lead to the effect of this value being ignored in the final loss function. Such an effect is also true for the Weibo dataset, so we finally chose a threshold $\eta$ value of \textbf{0.01}.

%--------------------------------------------------------------------------------4.8

\subsection{Case Study}
\begin{figure*}[!t]
    \centering
    \includegraphics[width=2\columnwidth]{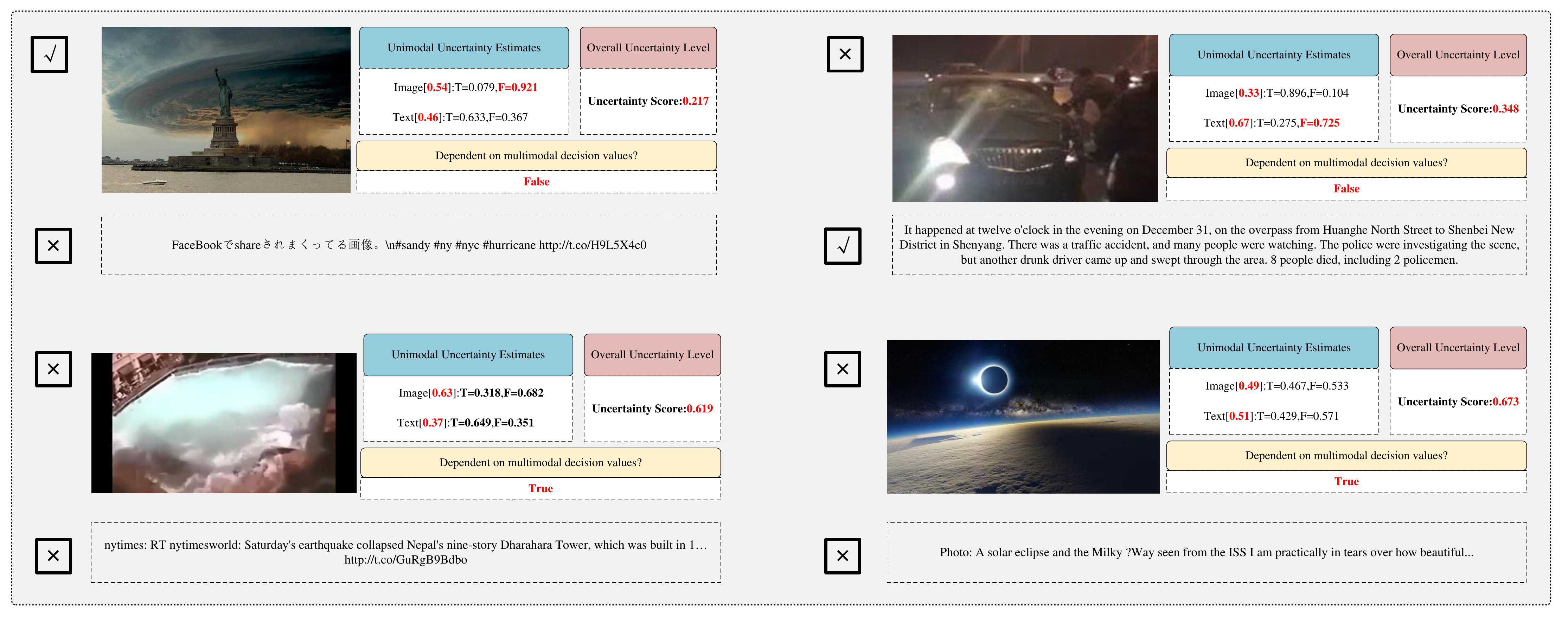}
    \caption{We show some examples of case studies for the dynamic fusion framework. T and F in the figure represent the prediction as true news and false news, respectively. the * in unimodal[*] represents the uncertainty score in the modality.}
    \label{fig:Casestudy}
\end{figure*}

In order to more deeply investigate the effectiveness of the dynamic fusion mechanism in the multimodal fake news detection task, we conducted a set of example case studies, as shown in Fig.\ref{fig:Casestudy}.We exemplify a total of four representative cases from the Twitter dataset and the Weibo dataset, respectively, and show in detail the overall uncertainty scores of the two modalities as well as the respective unimodal uncertainty scores.Note that the uncertainty score reflects how credible the modality is for the final prediction outcome,and it is inversely related to uncertainty.As shown in Algorithm \ref{alg:alg1}, when the value is high, especially when it is larger than the threshold we set, we directly use the decision opinion of the multimodal heterogeneous graph. And when this value is low we focus our attention on the decision results for unimodal text and unimodal images. In the first case, where the overall uncertainty is relatively low and there is a large difference between the uncertainty of the two unimodal modalities, we can clearly observe that the unimodal text has a lower uncertainty score, and therefore the MDF framework ends up adopting the decision opinion of the image. The opposite situation occurs in the second case, when the confidence level of the unimodal text is much higher than that of the unimodal image, and therefore we grant the final decision power to the unimodal text. When the overall uncertainty score is high, the decision-making power of the heterogeneous graph neural network is considered at this point, as shown in case four. Obtaining a plausible decision from only two unimodal uncertainty scores may produce poor results, e.g.,in the third case,the unimodal image has a much larger uncertainty score than the text, but the image itself is of low quality, so the data tends to have uncertainty. Fortunately, when overall uncertainty is taken into account, the final decision will no longer rely on a low quality unimodal image with a high uncertainty score, but rather the decision value of a multimodal heterogeneous image, which avoids catastrophic results to some extent.

%--------------------------------------------------------------------------------5

\section{Limitation}
\label{sec:limitation}
Although MDF has achieved excellent performance, especially in uncertainty modeling and dynamic fusion. However, we should also recognize that it has the following drawbacks:

$\bullet$ We assume that all tweets contain noise, which is unfair to the text of some normalized representations and to the data of images that have not been fabricated, especially for some official statements.

$\bullet$ The model suffers from the limitations of the D-S evidence theory, which is not very robust, and its soundness and validity are highly debatable, and in extreme cases the results derived from it are counter-intuitive, e.g., the "Zadeh Paradox".

$\bullet$ It is not reasonable to rely on tweets alone to make a decision, so our future work is to introduce external knowledge on top of it to improve the ability to detect fake news.

%--------------------------------------------------------------------------------6

\section{Conclusion}
\label{sec:conclusion}
Multimodal user posts containing fake news present on social media contain the problem of data uncertainty caused by unimodal noise. Therefore, we propose in this paper a dynamic fusion framework MDF for multimodal fake news detection, which can model inter- and intra-modal uncertainty while giving final decision results using a dynamic fusion strategy.The MDF framework first contains a UEM module for performing uncertainty modeling, which will map the unimodal text and the unimodal visual features through the multi-head attention mechanism mapping to a latent subspace that satisfies a Gaussian distribution. Then, a DFN module based on the graph attention mechanism with evidence theory is proposed to conveniently capture the inter-modal uncertainty information and is responsible for dynamically balancing the weights of the two modalities. Multiple sets of experiments on two publicly available datasets, Twitter and Weibo, demonstrate the effectiveness of the framework.

For further work, we will try to fully integrate more structured external knowledge and introduce external knowledge graphs to provide the model with more logical knowledge to further improve the ability to detect false news.

\section*{Ethical Consideration}
In this section, we check our paper with the following ethical considerations:

$\bullet$ \textbf{Avoid using Generative AI and AI-assisted technologies in the writing process}. We hereby declare that we have never used any form of generative artificial intelligence techniques and AI-assisted technologies in the writing of this manuscript to ensure that our content maintains the highest degree of originality, accuracy, and ethical responsibility.

$\bullet$ \textbf{Avoid harm}.This manuscript focuses on the design of a deep learning framework for multimodal fake news detection in which the participants were not harmed in any way.

$\bullet$ \textbf{Transparency}.We have given the source code of the design framework and the link to the repository of the data in the manuscript, which we will upload as soon as the paper is accepted. To ensure sufficient transparency of our work, we have given detailed experimental parameters as well as evaluation metrics in Section \ref{sec:4.2}, and analyzed the impact of the involved parameters on the final performance in Section \ref{sec:4.7}.

$\bullet$ \textbf{Public Good}.Our proposed MDF framework is only used to output information about the truthfulness of a certain multimodal news sample, which does not have any negative impact on social opinion.

$\bullet$\textbf{Confidentiality and Data Security}.We used two publicly available datasets, Twitter and Weibo, to evaluate our model, which have been used in many research works prior to this, and therefore do not have any data security issues or confidentiality concerns.

In summary, we assessed that our manuscript does not present any potential ethical risks in terms of harm, public good, Transparency, confidentiality and data security.
\section*{Acknowledgements}

The article is supported by the “Tianshan Talent” Research Project of Xinjiang (No.2022TSYCLJ0037).The National Natural Science Foundation of China (No.62262065). The Science and Technology Program of Xinjiang (No.2022 B01008).The National Key R\&D Program of China Major Project(NO.2022ZD0115800).

\bibliographystyle{cas-model2-names}

\bibliography{cas-refs}

\end{document}